\documentclass[12pt,draftcls,peerreviewca,onecolumn]{IEEEtran}

\usepackage{amsfonts}
\usepackage{cite}
\usepackage{graphicx}
\usepackage{amssymb}
\usepackage{subfigure}
\usepackage{array}
\usepackage{color}
\usepackage{amsmath}
\usepackage{algorithmic}
\usepackage{algorithm}
\usepackage{cases}
\usepackage{mathrsfs}
\usepackage{epsfig}
\usepackage{psfrag}
\usepackage{url}

\usepackage{etoolbox}
\makeatletter
\patchcmd{\@makecaption}
  {\scshape}
  {}
  {}
  {}
\makeatletter
\patchcmd{\@makecaption}
  {\\}
  {.\ }
  {}
  {}
\makeatother

\newcommand{\dis}{\stackrel{d}{\sim}}
\newcommand{\eqla}{\stackrel{(a)}{=}}
\newcommand{\eqlb}{\stackrel{(b)}{=}}
\newcommand{\eqlc}{\stackrel{(c)}{=}}
\newcommand{\eqld}{\stackrel{(d)}{=}}

\newtheorem{Thm}{Theorem}
\newtheorem{Lem}{Lemma}
\newtheorem{Cor}{Corollary}
\newtheorem{Def}{Definition}

\newtheorem{Prob}{Problem}
\newtheorem{Rem}{Remark}

\IEEEoverridecommandlockouts

\begin{document}

\title{Partition-based Caching  in Large-Scale SIC-Enabled Wireless Networks
}

\author{
\IEEEauthorblockN{Dongdong Jiang \thanks{D. Jiang and Y. Cui are with the Department of  Electronic Engineering, Shanghai Jiao Tong University, China. This paper will be  presented in part at IEEE ICC 2017, and is submitted in part to IEEE ISIT 2017.}} 
\and
\IEEEauthorblockN{Ying Cui}
}

\maketitle

\begin{abstract}
Existing designs for content dissemination do not fully explore and exploit potential caching and computation capabilities in advanced wireless networks.
In this paper, we propose two partition-based caching designs, i.e., a coded caching design based on Random Linear Network Coding and an uncoded caching design. We consider the analysis and optimization of the two caching designs in a large-scale successive interference cancelation (SIC)-enabled wireless network.
First, under each caching design, by utilizing tools from stochastic geometry and adopting appropriate approximations, we derive a tractable expression for the successful transmission probability in the general file size regime.
To further obtain design insights, we also derive closed-form expressions for the successful transmission probability in the small and large file size regimes, respectively.
Then, under each caching design, we consider the successful transmission probability maximization in the general file size regime, which is an NP-hard problem.
By exploring structural properties, we successfully transform the original optimization problem into a Multiple-Choice Knapsack Problem (MCKP), and obtain a near optimal solution with $\frac{1}{2}$ approximation guarantee and polynomial complexity. We also obtain closed-form asymptotically optimal solutions.
The analysis and optimization results show the advantage of the coded caching design over the uncoded caching design, and reveal the impact of caching and SIC capabilities. Finally, by numerical results, we show that the two proposed caching designs achieve significant performance gains over some baseline caching designs.
\end{abstract}


\newpage
\section{introduction}

The rapid proliferation of smart mobile devices has triggered an unprecedented growth of the global mobile data traffic. Recently,
caching at base stations (BSs) has been proposed as a promising  approach for  massive content delivery by reducing the distance between popular files and their requesters\cite{Bustag16magazine,Shanmugam13,cachingmimoLiu15}.
As the cache size is limited in general,  designing  caching strategies  appropriately is a prerequisite for  efficient content dissemination.
The performance of a caching design is highly affected by the file diversity it provides.
In \cite{EURASIP15Debbah,Bharath16TCOM,Cui16TWC,Cui16Hetnet}, the authors consider large-scale wireless network models where the stochastic nature of geographic locations of BSs and users is characterized using stochastic geometry.
In particular, in \cite{EURASIP15Debbah}, the authors  consider caching the most popular files at each BS, which does not provide file diversity.
In \cite{Bharath16TCOM}, the authors consider random caching with files being stored at each BS in an i.i.d. manner, which may store multiple copies of a file at one BS and yield storage waste.
In \cite{Cui16TWC,Cui16Hetnet}, the authors consider random caching  and multicasting on the basis of file combinations consisting of different files, and analyze and optimize the joint design.
Note that
the random caching designs in \cite{Bharath16TCOM,Cui16TWC,Cui16Hetnet} can provide file diversity.
However, in \cite{Bharath16TCOM,Cui16TWC,Cui16Hetnet}, a file transmission may not make full use of the file diversity provided by the random caching designs, when the serving BS of the file is not close to the file requester.
In addition, the caching designs  proposed in \cite{EURASIP15Debbah,Bharath16TCOM,Cui16TWC,Cui16Hetnet} require storing entire files at each BS, which may restrict file diversity and limit the potential of caching.

To further improve file diversity, in \cite{Gu13PIMRC,Altman13,BioglioGC15,Golrezaei12ICC,QuekTWC16},  files are partitioned into multiple subfiles, and each BS stores an uncoded or coded subfile of a file.
For instance, in \cite{Gu13PIMRC} and \cite{Altman13}, the authors propose network coding-based caching designs, and analyze the cache miss probability \cite{Gu13PIMRC} and minimize the occupied storage space \cite{Altman13}, respectively.
In \cite{BioglioGC15} and \cite{Golrezaei12ICC}, the authors propose MDS code-based caching designs, and minimize the backhaul rate \cite{BioglioGC15} and average delay \cite{Golrezaei12ICC}, respectively.
In \cite{QuekTWC16}, the authors propose a partition-based uncoded caching design, and analyze the successful content delivery probability.
Note that
the network coding-based caching designs in \cite{Gu13PIMRC,Altman13} are restricted to a single file and cannot be directly applied to practical networks with multiple files.
In addition, the coded caching designs in \cite{Gu13PIMRC,Altman13,BioglioGC15,Golrezaei12ICC} do not consider the delivery of the cached files, and hence may not yield good user experience.
Compared to coded caching designs, the uncoded caching design in \cite{QuekTWC16} may not sufficiently exploit storage resources. But \cite{QuekTWC16} considers the delivery of the cached files, where successive interference cancelation (SIC) is employed at each user to decode all the uncoded subfiles of its desired file transmitted at the same time over the same frequency band. 
Given computational capability at users, applying SIC can facilitate the exploitation of file diversity, and hence improve the performance of the caching design.

SIC is a promising technique to improve the performance of wireless networks with relatively small additional computational complexity. The idea of SIC is to decode multiple signals sequentially by subtracting interference due to the decoded signals before decoding other signals. The use of SIC hinges on the imbalance of the received powers of different signals, which come from transmitters at different locations in some cases.
Conventional performance analysis of SIC is for wireless networks with transmitters residing at given locations \cite{Zanella12TCOM}.
To capture the impact of the spatial distribution of transmitters, recent studies attempt to reveal the gain of SIC in large-scale wireless networks utilizing tools from stochastic geometry \cite{Weber07TIT,Zhang14TIT,Quek14TCOM,Ma15INFOCOM}.
In this context, approximations are often used due to the well-acknowledged challenge in tracking the problem directly.
For instance, in \cite{Weber07TIT}, the authors characterize the performance of SIC using a guard-zone based approximation, where the interferers inside a guard-zone centered at the receiver are assumed canceled, and the size of the guard-zone is used to model the SIC capability.
In \cite{Zhang14TIT}, the authors consider SIC based on power order, i.e., from the stronger signals to the weaker signals, and derive tractable bounds on the successful decoding probability.
In \cite{Quek14TCOM} and \cite{Ma15INFOCOM}, the authors consider SIC based on distance order, i.e., from the nearer transmitters to the farther transmitters, and obtain closed-form expressions for the coverage probabilities in heterogeneous networks and D2D networks, respectively, assuming independence between decoding events.
Note that \cite{Zanella12TCOM,Weber07TIT,Zhang14TIT,Quek14TCOM,Ma15INFOCOM} focus on performance analysis of SIC in large-scale wireless networks without caching capability.

In summary, further studies are required to understand the fundamental impacts of communication, caching and computation (e.g., SIC) capabilities on network performance.
In this paper, we shall  shed some light on the essential problem. We consider multiple files and 
a large-scale SIC-enabled wireless network with random channel fading as well as stochastic locations of BSs and users.
Our main contributions are summarized below.
\begin{itemize}
\item First, we propose two general partition-based caching designs, i.e., a coded caching design based on Random Linear Network Coding (RLNC)  \cite{Medard06TIT} and an uncoded caching design, which incorporate any deterministic and identical (same at all BSs) caching of entire files as a special case. Correspondingly, we transmit multiple coded or uncoded subfiles of a requested file at the same time over the same frequency band, and adopt SIC to decode these subfiles for recovering the requested file.

\item Then, we analyze the successful transmission probability. The challenge in analyzing partition-based caching and SIC is commonly recognized. By utilizing tools from stochastic geometry and adopting appropriate approximations, under each caching design, we derive a tractable expression for the successful transmission probability in the general file size regime. We also show that the coded caching design outperforms the uncoded caching design in the general file size regime. To further obtain design insights, under each caching design, we derive closed-form expressions for the successful transmission probabilities in the small and large file size regimes, respectively, utilizing series expansion of some special functions. These expressions reveal the impacts of caching and SIC capabilities.
    From the asymptotic analysis, we know that under each caching design, the successful transmission probability increases linearly as the file size decreases to zero, and decreases exponentially to zero as the file size increases to infinity.

\item Next, we consider the successful transmission probability maximization by optimizing a design parameter. In the general file size regime, under each caching design, by exploring structural properties of the optimization problem which is NP-hard, we successfully transform the original optimization problem into a Multiple-Choice Knapsack Problem (MCKP), and obtain a near optimal solution with $\frac{1}{2}$ approximation guarantee and polynomial complexity \cite{book04Hans}.
    Under the coded caching design, we also obtain closed-form asymptotically optimal solutions in the small and large file size regimes, respectively. Under the uncoded caching design, we obtain a near optimal solution in the small file size regime and a closed-form asymptotically optimal solution in the large file size regime, respectively.
    From the asymptotic optimization results, we know that in the small file size regime, the optimal successful transmission probability under the coded caching design increases with the cache size and the SIC capability. While, in the large file size regime, the optimal successful transmission probabilities under both caching designs increase with the cache size and are not affected by the SIC capability.
    The asymptotic optimization results also demonstrate the advantage of the coded caching design over any deterministic and identical caching of  entire files in the small file size regime.

\item Finally, by numerical results, we show that the proposed coded caching design achieves a significant performance gain in the successful transmission probability in the general file size regime over the proposed uncoded caching design and some baseline caching designs.
\end{itemize}

%
%

%
%
%
%

\section{system model and performance metric}

\begin{figure}[t]
\begin{center}
 \includegraphics[width=13cm]{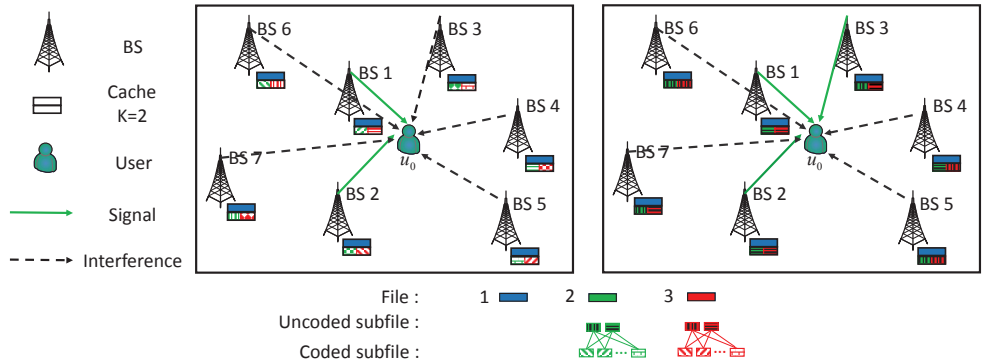}
  \end{center}
  \caption{\small{System model. $N=3$, $K=2$, $s_1=1$, $s_2=s_3=\frac{1}{2}$, $m_1=1$ and $m_2=m_3=3$.
  }
 }
\label{fig:system}
\end{figure}

\subsection{Network Model}\label{subsec:netmodel}
We consider a  large-scale wireless  network, as shown in Fig. \ref{fig:system}. The locations of the base stations (BSs) are spatially distributed as a two-dimensional homogeneous Poisson point process (PPP) $\Phi_{b}$ with density $\lambda_{b}$. We focus on a typical user $u_0$, which we assume without loss of generality (w.l.o.g.) to be located at the origin.
The BSs are labeled  in ascending order of distance from $u_0$. 
Let $d_i$ denote the distance between  BS $i\in \Phi_b$ and $u_0$. Thus, we have  $d_1\leq d_2\leq\cdots$.
We consider the downlink transmission. Each BS has one transmit antenna and transmits with power $P$ over bandwidth $W$. User $u_0$ has one receive antenna. Consider a discrete-time system with time being slotted. The duration of each time slot is $T$ seconds. We study one slot of the network. We consider both path loss and small-scale fading. Specifically, due to path loss,  transmitted signals with distance $d$ are attenuated by a factor $d^{-\alpha}$, where $\alpha>2$ is the path loss exponent. For small-scale fading, we assume Rayleigh fading, i.e., each small-scale channel  $h\dis\mathcal {CN}(0, 1)$.

Let $\mathcal N\triangleq \{1,2,\cdots, N\}$ denote the set of $N$ files in the network. For ease of illustration, we assume that all  files  have the same size of $S$ bits. Each file is of certain popularity. User $u_0$ randomly  requests one file, which is file $n\in \mathcal N$ with probability $a_n\in (0,1)$, where $\sum_{n\in \mathcal N}a_n=1$.  Thus, the file popularity distribution is given by $\mathbf a\triangleq (a_n)_{n\in \mathcal N }$, which   is assumed to be known apriori.  In addition, w.l.o.g., we assume $a_{1}> a_{2}\ldots> a_{N}$.

\subsection{Caching Designs}\label{subsec:description}
The  network consists of  cache-enabled BSs. In particular, each BS is equipped with a cache of size $K\geq 1$ (in files), i.e., $KS$ (in bits).
Assume each BS cannot store all files in $\mathcal N$ due to the limited storage capacity, i.e., $K<N$.
Now, we propose two caching designs, i.e., a coded caching design based on Random Linear Network Coding (referred to as RLNC caching design)\cite{Medard06TIT} and  an uncoded caching design (referred to as UC caching design), as illustrated in Fig.~\ref{fig:system}. Both are partition-based and are parameterized by $\mathbf s\triangleq\left(s_n\right)_{n\in\mathcal N}$, where $s_n\in\{0\}\cup \left\{\frac{1}{m}\Big{|}m\in\mathbb N^+\right\},\; n\in\mathcal N$
represents the amount of storage (in files) allocated to file $n\in\mathcal N$ at each BS. Here, $\mathbb N^+$ denotes the set of positive integers.
In particular, for any file $n\in\mathcal N$, consider the following three cases. (i) If $s_n=0$, file $n$ is not stored at any BS. (ii) If $s_n=1$, file $n$ is stored at each BS. (iii) If $s_n\in\{\frac{1}{m}|m=2,3,\cdots\}$, file $n$ is partitioned into $\frac{1}{s_n}$ subfiles, each of $s_nS$ bits.
In Case (i) and Case (ii), the two caching designs coincide. In Case (iii), the two caching designs differentiate with each other as follows.
\begin{itemize}
\item {\em RLNC Caching Design in Case (iii)}:
Each BS forms a random linear combination of all the $\frac{1}{s_n}$ subfiles of file $n$ (i.e., a coded subfile of file $n$ which is of $s_nS$ bits) using RLNC, and stores it in its cache.
We consider RLNC over a large field, and assume that file $n$ can be decoded from any $\frac{1}{s_n}$ coded subfiles of file $n$ stored in the network \cite{Medard06TIT}.

\item {\em UC Caching Design in Case (iii)}:
Each BS selects a random subfile from the $\frac{1}{s_n}$ subfiles of file $n$  according to the uniform distribution, and stores it in its cache.
\end{itemize}
The design parameter $\mathbf s$ of a feasible RLNC or UC caching design satisfies the following constraint
\small{\begin{align}
\sum_{n\in\mathcal N}s_n\leq K. \label{eqn:cache_size_constr_coding}
\end{align}}\normalsize
\begin{Rem}[Relation with Deterministic and Identical Caching of Entire Files]\label{Rem:relation_between_caching}
If $s_n\in\{0,1\}$ for all $n \in \mathcal N$, the two proposed caching designs degenerate to deterministic and identical caching of entire files, a typical example of which is caching the most popular (entire) files at each BS. Thus, the two proposed partition-based caching designs are more general.
\end{Rem}

\subsection{File Transmission and Reception}\label{subsec:file_transmission_and_reception}

Assume that each BS $i$ knows that it is the $i$th nearest BS of $u_0$, and is not aware of the file placement of other BSs.
We now introduce the file transmission under the two proposed caching designs, as illustrated in Fig. \ref{fig:system}.
The file transmission under the RLNC caching design is determined by parameter $\mathbf s$ only. While, the file transmission strategy under the UC caching design depends on parameter $\mathbf s$ and another parameter $\mathbf m\triangleq (m_n)_{n\in\mathcal N}$, where $m_n\in\{\frac{1}{s_n},\frac{1}{s_n}+1,\cdots\}$ represents the number of (nearest) BSs serving the request for file $n$.
Specifically, consider the following three cases.
(i) If $s_n=0$, $u_0$ cannot obtain file $n$ from a cache of the network under either caching design.\footnote{In this case, $u_0$ may be served through other service mechanisms. For example, BSs can fetch some uncached files from the core network through  backhaul links and transmit  them  over other reserved frequency bands. The service of uncached files may involve  backhaul cost or extra delay. The investigation of service mechanisms for uncached files  is beyond the scope of this paper.} In this case, we set $m_n=0$ for the UC cachign design.
(ii) If $s_n=1$, the nearest BS transmits file $n$ to $u_0$ under both caching designs. In this case, we set $m_n=1$ for the UC cachign design.
(iii) If $s_n\in\{\frac{1}{m}|m=2,3,\cdots\}$, each of the $\frac{1}{s_n}$ nearest BSs transmits the stored coded subfile of file $n$ to $u_0$ under the RLNC caching design, and each of the $m_n$ nearest BSs transmits the stored uncoded subfile of file $n$ to $u_0$ under the UC caching design.
Note that in Cases (ii) and (iii), the subfile transmission of each serving BS is over the whole bandwidth and time slot.

In Case (iii), for the RLNC caching design, file $n$ can be decoded from the $\frac{1}{s_n}$ coded subfiles of file $n$ stored in the $\frac{1}{s_n}$ nearest BSs; for the UC caching design, file $n$ may not be successfully decoded from the $m_n$ uncoded subfiles of file $n$ stored in the $m_n$ nearest BSs, as the $m_n$ subfiles may not cover the $\frac{1}{s_n}$ different subfiles (due to the random placement of the $\frac{1}{s_n}$ different subfiles under the UC caching design).
For the UC caching design, let $I_n^j\in\mathbb N^+$ denote the nearest BS storing the $j$th subfile of file $n$, where $j\in\{1,2,\cdots,\frac{1}{s_n}\}$, and denote $I_n^u\triangleq \max\big\{I_n^1,I_n^2,\cdots,I_n^{\frac{1}{s_n}}\big\}\in\{\frac{1}{s_n},\frac{1}{s_n}+1,\cdots\}$. Note that $I_n^j$, $j\in\{1,2,\cdots, \frac{1}{s_n}\}$ and $I_n^u$ are random variables with the randomness induced by the random subfile placement, and the probability mass functions (p.m.f.s) of $I_n^j$, $j\in\{1,2,\cdots,\frac{1}{s_n}\}$ and $I_n^u$ depend on $s_n$.
In particular, if $I_n^u\leq m_n$, file $n$ can be decoded from the subfiles of file $n$ stored in the $m_n$ nearest BSs; otherwise file $n$ cannot be decoded.
For ease of illustration, we also set $I_n^u=\infty$ in Case (i) (i.e., $s_n=0$) and $I_n^u=1$ in Case (ii) (i.e., $s_n=1$).
%

In this paper, we consider an interference-limited network and neglect the background thermal noise. We assume all BSs are active for serving their own users.
Thus, the received signals of $u_0$ under the RLNC caching design and the UC caching design, denoted as $y_n^c$ and $y_n^u$ respectively, are given by
\small{\begin{align}
y_n^c&=\sum_{i\in \mathbf \{1,2,\cdots,\frac{1}{s_n}\}}d_i^{-\frac{\alpha}{2}}h_ix_i+\sum_{i\in  \Phi_b\setminus \{1,2,\cdots,\frac{1}{s_n}\}}d_i^{-\frac{\alpha}{2}}h_ix_i\;,\label{eqn:receive_signal_1st_strategy}\\
y_n^u&=\sum_{i\in\{1,2,\cdots,m_n\}}d_i^{-\frac{\alpha}{2}}h_ix_i+\sum_{i\in  \Phi_b\setminus \{1,2,\cdots,m_n\}}d_i^{-\frac{\alpha}{2}}h_ix_i,\label{eqn:receive_signal_partition}
\end{align}}\normalsize
where $d_i$ is the distance between BS $i$ and $u_0$, $h_i\dis\mathcal {CN}(0, 1)$ is the small-scale channel between BS $i$ and $u_0$, $x_i$ is the transmit signal  from BS $i$. The first sums in \eqref{eqn:receive_signal_1st_strategy} and \eqref{eqn:receive_signal_partition} represent the desired signals, and the second sums in \eqref{eqn:receive_signal_1st_strategy} and \eqref{eqn:receive_signal_partition} represent the interferences.

To extract multiple signals under each caching design, we adopt SIC.
As in \cite{Quek14TCOM}, we consider the distance-based decoding and cancelation order.\footnote{It has been shown that the order statistics of received powers are dominated by distances\cite{Quek14TCOM}.}
In particular, when decoding the signal from BS $i$, all signals from the nearer BSs in $\{1,2,\cdots,i-1\}$ need to be successfully decoded and canceled, where $i\in\{1,2,\cdots,\frac{1}{s_n}\}$ for the RLNC caching design and $i\in\{1,2,\cdots,I_n^u\}$ for the UC caching design.\footnote{In distance-based decoding and cancelation order, under the UC caching design, the signals from the $I_n^u$ nearest BSs have to be decoded, in order to obtain $\frac{1}{s_n}$ different signals for recovering file $n$.}
The signal-to-interference ratio (SIR) of the signal from BS $i$ after successfully decoding and canceling the signals from the nearer BSs in $\{1,2,\cdots,i-1\}$ is given by
\small{\begin{align}
{\rm SIR}_i=\frac{d_i^{-\alpha}|h_i|^2}{I_i},\label{eqn:SIR_m_expression}
\end{align}}\normalsize
where $I_i\triangleq \sum_{j\in\Phi_b\setminus \{1,2,\cdots,i\}}d_j^{-\alpha}|h_j|^2$ denotes the interference in decoding the signal from BS $i$.
If $W{\rm log}_2(1+{\rm SIR}_i)>\frac{s_nS}{T}$, $u_0$ can successfully decode (and cancel) the signal from BS $i$.
Due to the limited computational capability and the delay constraint, as in \cite{Quek14TCOM}, we assume that $u_0$ has limited SIC capability $M$, which is a system parameter. That is, $u_0$ can perform decoding and cancelation at most $M$ times to obtain its desired signals.
Denote $\mathcal S\triangleq\{0\}\cup\{\frac{1}{m}|m=1,2,\cdots,M\}$ and $\mathcal U \triangleq \{(0,0)\}\cup\{(1,1)\}\cup\big{\{}(s,m)|s\in\mathcal S\setminus\{0,1\}, m\in\{\frac{1}{s},\frac{1}{s}+1,\cdots,M\}\big{\}}$.
Given SIC capability $M$, for the two proposed caching designs to be meanful, we require
\small{\begin{align}
&s_n\in\mathcal S, \quad n\in\mathcal N,\label{eqn:division_choise_req}\\
&(s_n,m_n)\in\mathcal U,\quad n\in\mathcal N,\label{eqn:constr_UC}
\end{align}}\normalsize
where \eqref{eqn:division_choise_req} is for the RLNC caching design, and \eqref{eqn:constr_UC} is for the UC caching design.
Note that when $M=1$, \eqref{eqn:division_choise_req} and \eqref{eqn:constr_UC} reduce to $s_n\in\{0,1\}$, $n\in\mathcal N$ and $(s_n,m_n)\in\{(0,0),(1,1)\}$, $n\in\mathcal N$, respectively. Thus, when $M=1$, the two proposed caching designs degenerate to deterministic and identical caching of entire files.

\subsection{Performance Metric}

Requesters are mostly concerned about whether  their desired files can be successfully received. Therefore, in this paper, we consider the successful transmission probability  of a file randomly requested by $u_0$ as the network performance metric.
According to the file transmission and reception discussed in Section \ref{subsec:file_transmission_and_reception}, the successful transmission probabilities of file $n\in\mathcal N$ requested by $u_0$ under the RLNC caching design and the UC caching design, denoted as $q_n^c(s_n)$ and $q_n^u(s_n,m_n)$ respectively, are given by
\small{\begin{align}
&q_n^c(s_n)=
\begin{cases}
0,& s_n=0\\
\Pr\left[\bigcap\limits_{i\in\{1,2,\cdots,\frac{1}{s_n}\}}W{\rm log}_2(1+{\rm SIR}_i)>\frac{s_nS}{T}\right],\quad &s_n\in\mathcal S\setminus\{0\}
\end{cases},\label{eqn:q_n^c_orig}\\
&q_n^u(s_n,m_n)\notag\\
&=\begin{cases}
0,&(s_n,m_n)=(0,0)\\
\Pr\left[I_n^u\in\{\frac{1}{s_n},\frac{1}{s_n}+1,\cdots,m_n\},\;\bigcap\limits_{i\in\{1,2,\cdots,I_n^u\}}W{\rm log}_2(1+{\rm SIR}_i)>\frac{s_nS}{T}\right],\ & (s_n,m_n)\in\mathcal U\setminus\{(0,0)\}
\end{cases},\label{eqn:q_n^u_orig}
\end{align}}\normalsize
where ${\rm SIR}_i$ is given by \eqref{eqn:SIR_m_expression}.
According to the total probability theorem, the successful transmission probabilities of a file randomly requested by $u_0$ under the RLNC caching design and the UC caching design, denoted as $q^c(\mathbf s)$ and $q^u(\mathbf s,\mathbf m)$ respectively, are given by\footnote{Here, $\mathcal X^k$ denotes the $k$-ary Cartesian power of set $\mathcal X$. 
}
\small{\begin{align}
&q^c(\mathbf s)=\sum_{n\in\mathcal N}a_nq_n^c(s_n), \quad \mathbf s\in\mathcal S^{N},\notag\\
&q^u(\mathbf s,\mathbf m)=\sum_{n\in\mathcal N}a_nq_n^u(s_n,m_n), \quad (\mathbf s,\mathbf m)\in\mathcal U^{N}.\notag
\end{align}}\normalsize

\section{Performance Analysis and Optimization of RLNC caching design}

In this section, we consider the performance analysis and optimization of the RLNC caching design. First, we analyze the successful transmission probabilities in the general file size regime, the small file size regime and the large file size regime, respectively. Then, we optimize the successful transmission probabilities in these regions.

\subsection{Performance Analysis of RLNC caching Design}

\subsubsection{Performance Analysis in General File Size Regime}

First, we calculate the probability of successfully decoding the $i$ subfiles of file $n$, each of $s_nS$ bits, stored at the $i$ nearest BSs, i.e., $h(s_n,i)\triangleq \Pr\left[\bigcap_{j\in\{1,2,\cdots,i\}}W{\rm log}_2(1+{\rm SIR}_j)>\frac{s_nS}{T}\right]$, where $s_n\in\mathcal S\setminus\{0\}$.
The calculation of $h(s_n,i)$ requires the conditional joint probability density function (p.d.f.) of $d_1^{-\alpha}|h_1|^2, I_1, I_2,\cdots, I_{i}$ conditioned on the distances $d_1,d_2,\cdots,d_{i}$, which is difficult to obtain.
As in \cite{Quek14TCOM}, for the tractability of the analysis, we assume the independence between the events $W{\rm log}(1+{\rm SIR}_j)>\frac{s_nS}{T},\; j\in\{1,2,\cdots,i\}$. Then, we have the following lemma.
\begin{Lem}\label{Lem:exp_of_h_s_i}
We have
\small{\begin{align}
h(s_n,i)=\frac{1}{\left(1+\frac{2}{\alpha}(2^{\frac{s_nS}{TW}}-1)^{\frac{2}{\alpha}}
B'\left(\frac{2}{\alpha},1-\frac{2}{\alpha},2^{-\frac{s_nS}{TW}}\right)\right)^{\frac{i(i+1)}{2}}},\quad s_n\in\mathcal S\setminus\{0\},\label{eqn:h_z_i}
\end{align}}\normalsize
where $B'(x,y,z)\triangleq \int_z^1 u^{x-1}(1-u)^{y-1}{\rm d}u$ is the complementary incomplete Beta function.
\end{Lem}
\begin{IEEEproof}
Please refer to Appendix A.
\end{IEEEproof}

From \eqref{eqn:q_n^c_orig}, we have
\small{\begin{align}
q_n^c(s_n)=
\begin{cases}
0, &s_n=0\\
h\left(s_n,\frac{1}{s_n}\right), \quad &s_n\in\mathcal S\setminus\{0\}
\end{cases}.\label{eqn:q_n^c}
\end{align}}\normalsize
Therefore, we have the successful transmission probability under the RLNC caching design, as summarized below.
\begin{Thm}[Performance of RLNC in General File Size Regime]\label{Thm:CP_close_form_dist_order}
The successful transmission probability $q^c(\mathbf s)$ under the RLNC caching design is given by
\small{\begin{align}
q^c(\mathbf s)&=\sum_{n\in\mathcal N}a_nq_n^c(s_n), \quad \mathbf s\in\mathcal S^{N},\label{eqn:CP_coding_caching_close_form}
\end{align}}\normalsize
where $q_n^c(s_n)$ is given by \eqref{eqn:q_n^c}.
\end{Thm}

\begin{figure}[t]
\begin{center}
\includegraphics[width=6cm]{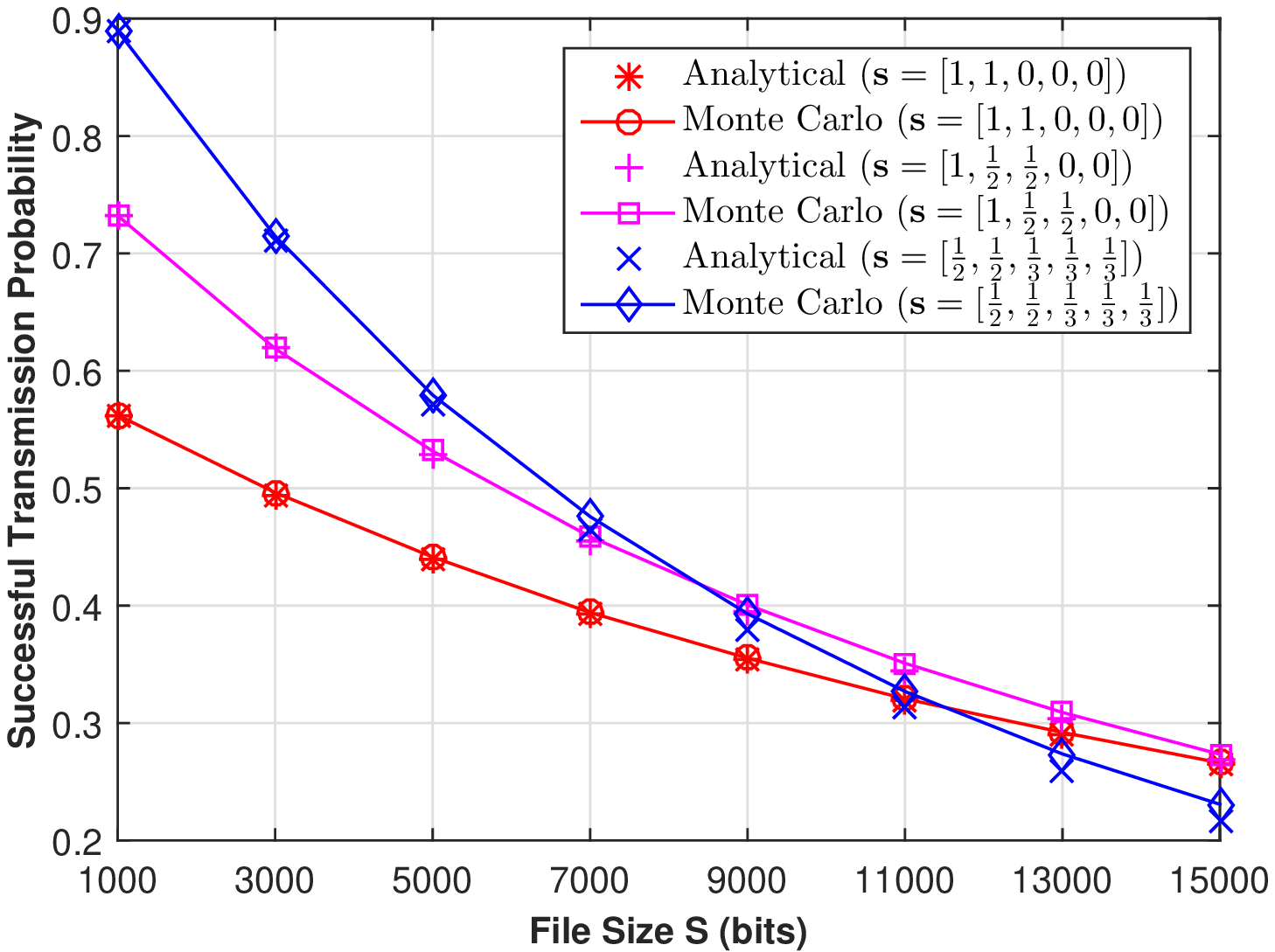}
\end{center}
\caption{Successful transmission probability under the RLNC caching design versus file size $S$. $N=5$, $K=2$, $\alpha=4$, $W = 10$MHz, $T = 1$ms, $M = 3$, and $a_n=\frac{n^{-\gamma}}{\sum_{n\in \mathcal N}n^{-\gamma}}$ with $\gamma=1$.}
\label{fig:verification_anal_MC}
\end{figure}

From Theorem \ref{Thm:CP_close_form_dist_order}, we can see that  $q^c(\mathbf s)$ is a decreasing function of $S$. The impact  of $\mathbf s$ on $q^c(\mathbf s)$ is not obvious.
Fig. \ref{fig:verification_anal_MC} plots $q^c(\mathbf s)$ versus $S$ at different $\mathbf s$. Fig. \ref{fig:verification_anal_MC} verifies Theorem \ref{Thm:CP_close_form_dist_order} and demonstrates the accuracy of the approximation adopted.
In addition, from Fig. \ref{fig:verification_anal_MC}, we can see that $q^c(\mathbf s)$ decreases with $S$. The impact of $\mathbf s$ on $q^c(\mathbf s)$ is not clear in the general file size regime.
To further obtain design insights, in the following,  we analyze the asymptotic  successful transmission probabilities in the small file size regime and the large file size regime, respectively.


\subsubsection{Performance Analysis in Small File Size Regime}

Utilizing series expansion of some special functions, from  Theorem \ref{Thm:CP_close_form_dist_order}, we derive the asymptotic successful transmission probability in the  small file size regime (i.e., $S\to 0$) as follows.
\begin{Lem}[Performance of RLNC in Small File Size Regime]\label{Lem:CP_asymp_0}
For all $\mathbf s\in\mathcal S^{N}$, we have $q^c(\mathbf s)\stackrel{S\to0}{\sim}q_{0}^c(\mathbf s)$,\footnote{$f(S)\stackrel{S\to0}{\sim}g(S)$ means $\lim_{S\to 0}\frac{f(S)}{g(S)}=1$.} where
\small{\begin{align}
q_0^c(\mathbf s)\triangleq \sum_{n\in\mathcal N}a_n\mathbf 1\left[s_n\neq 0\right]-\frac{{\rm ln}2}{(\alpha-2)WT}S\sum_{n\in\mathcal N}a_n\left(1+\frac{1}{s_n}\right)\mathbf 1\left[s_n\neq 0\right].\label{eqn:CP_coding_caching_aympt_0}
\end{align}}\normalsize
Here, $\mathbf 1[\cdot]$ denotes the indicator function.
\end{Lem}
\begin{IEEEproof}
Please refer to Appendix B.
\end{IEEEproof}

\begin{figure}[t]
\begin{center}
 \subfigure[\small{Small file size regime.}]
 {\resizebox{6cm}{!}{\includegraphics{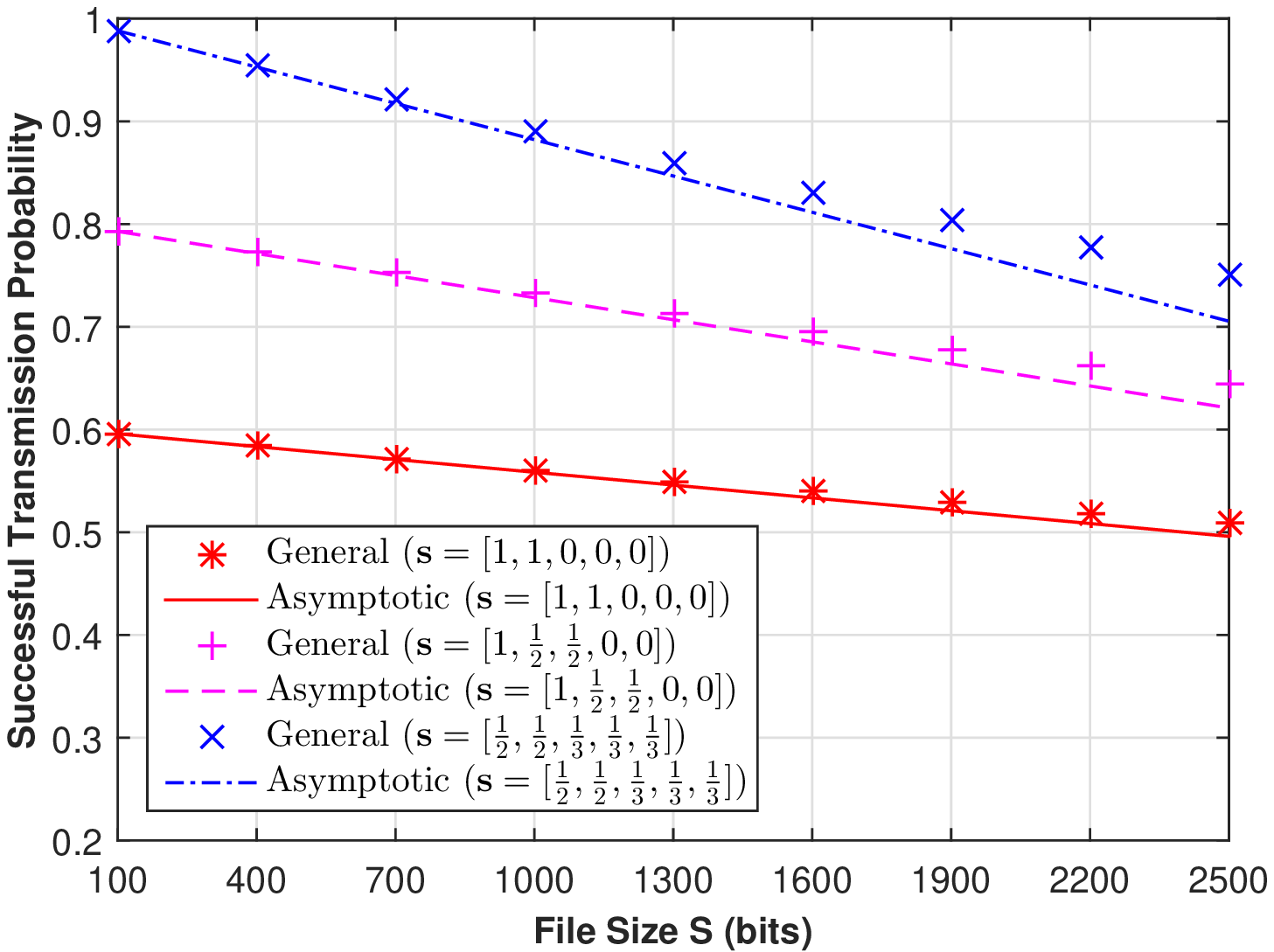}}}\quad\quad
 \subfigure[\small{Large file size regime.}]
 {\resizebox{6cm}{!}{\includegraphics{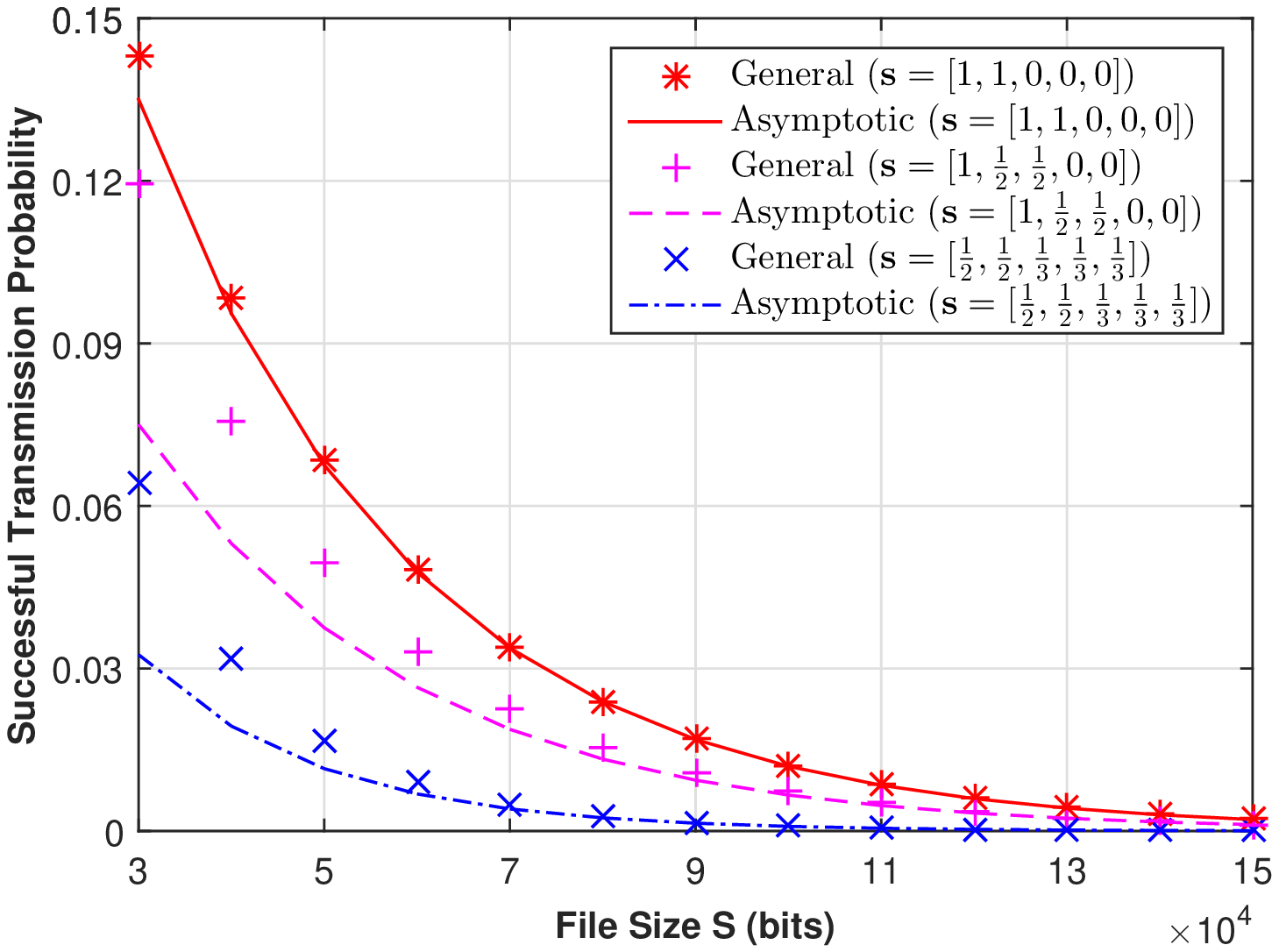}}}
 \end{center}
   \caption{\small{Successful transmission probability under the RLNC caching design versus file size $S$.  $N=5$, $K=2$, $\alpha=4$, $W = 10$MHz, $T = 1$ms, $M = 3$,  and $a_n=\frac{n^{-\gamma}}{\sum_{n\in \mathcal N}n^{-\gamma}}$ with $\gamma=1$.}}
\label{fig:asymp_STP}
\end{figure}

From Lemma \ref{Lem:CP_asymp_0}, we know that $\lim_{S\to 0}q^c(\mathbf s)=\sum_{n\in\mathcal N}a_n\mathbf 1\left[s_n\neq 0\right]$, and $q_0^c(\mathbf s)$ increases linearly to $\sum_{n\in\mathcal N}a_n\mathbf 1\left[s_n\neq 0\right]$ as $S$ decreases to $0$.
In addition, $\mathbf s$  affects the asymptotic behavior of $q_0^c(\mathbf s)$ by affecting $\lim_{S\to 0}q^c(\mathbf s)$ and the coefficient of $S$.
Fig.~\ref{fig:asymp_STP}~(a) plots $q_0^c(\mathbf s)$ versus $S$ in the small file size regime. We see from Fig.~\ref{fig:asymp_STP}~(a) that when $S$ decreases, the gap between each ``General" curve, which is plotted using Theorem~\ref{Thm:CP_close_form_dist_order}, and  the corresponding ``Asymptotic" curve, which is plotted  using Lemma~\ref{Lem:CP_asymp_0}, decreases. Thus, Fig.~\ref{fig:asymp_STP}~(a) verifies Lemma~\ref{Lem:CP_asymp_0}.

\subsubsection{Performance Analysis in Large File Size Regime}
Utilizing series expansion of some special functions, from Theorem \ref{Thm:CP_close_form_dist_order}, we derive the asymptotic successful transmission probability in the large file size regime (i.e., $S\to \infty$) as follows.
\begin{Lem}[Performance of RLNC in Large File Size Regime]\label{Lem:CP_asymp_CP}
For all $\mathbf s\in\mathcal S^{N}$, we have $q^c(\mathbf s)\stackrel{S\to\infty}{\sim}q_{\infty}^c(\mathbf s)$,\footnote{$f(S)\stackrel{S\to\infty}{\sim}g(S)$ means $\lim_{S\to \infty}\frac{f(S)}{g(S)}=1$.} where
\small{\begin{align}
q_{\infty}^c(\mathbf s)&\triangleq \frac{2^{-\frac{\left(s_{\rm max}+1\right)S}{\alpha s_{\rm max}WT}}}{\left(\frac{2}{\alpha}
B\left(\frac{2}{\alpha},1-\frac{2}{\alpha}\right)\right)^{\frac{s_{\rm max}+1}{2s_{\rm max}^2}}}\sum_{n\in\mathcal N}a_n\mathbf 1\left[s_n=s_{\rm max}\right].\label{eqn:CP_coding_caching_aympt}
\end{align}}\normalsize
Here, $s_{\rm max}\triangleq \max\{s_n|n\in\mathcal N\}$
and $B(x,y)\triangleq\int_0^1u^{x-1}(1-u)^{y-1}{\rm d}u$ is the Beta function.
\end{Lem}
\begin{IEEEproof}
Please refer to Appendix C.
\end{IEEEproof}

From Lemma \ref{Lem:CP_asymp_CP}, we know that $\lim_{S\to \infty}q^c(\mathbf s)=0$, and $q_{\infty}^c(\mathbf s)$ decreases exponentially to $0$ as $S$ increases to $\infty$.
In addition, $\mathbf s$  affects the asymptotic behavior of $q_{\infty}^c(\mathbf s)$ via $s_{\rm max}$ only.
Fig.~\ref{fig:asymp_STP}~(b) plots $q_{\infty}^c(\mathbf s)$ versus $S$ in the large file size regime. We see from Fig.~\ref{fig:asymp_STP}~(b) that when $S$ increases, the gap between each ``General" curve, which is plotted using Theorem~\ref{Thm:CP_close_form_dist_order}, and the corresponding ``Asymptotic" curve, which is plotted using Lemma~\ref{Lem:CP_asymp_CP}, decreases. Thus, Fig.~\ref{fig:asymp_STP}~(b) verifies Lemma~\ref{Lem:CP_asymp_CP}.

\subsection{Performance Optimization of RLNC Caching Design}


\subsubsection{Performance Optimization in General File Size Regime}\label{sec:RLNC_opt_general}
The caching design affects the successful transmission probability $q^c(\mathbf s)$ via design parameter $\mathbf s$. We would like to maximize $q^c(\mathbf s)$ by carefully optimizing $\mathbf s$.
\begin{Prob}[RLNC Caching Design in General File Size Regime]\label{prob:opt_coding}
\small{\begin{align}
q^{c*}\triangleq \max_{\mathbf{s}} &\quad  q^c(\mathbf s)\notag\\
s.t. &\quad\eqref{eqn:cache_size_constr_coding}, \eqref{eqn:division_choise_req},\notag
\end{align}}\normalsize
where $q^c(\mathbf s)$ is given by \eqref{eqn:CP_coding_caching_close_form}. Let $\mathbf s^*$ denote the optimal solution.
\end{Prob}

When $M=1$, we have $s_n\in\{0,1\}$ for all $n\in\mathcal N$, and the proposed RLNC caching design degenerates to deterministic and identical caching of entire files. Based on structural properties of $q^{c}(\mathbf s)$, we know that the optimal caching design is to store the most $K$ popular entire files  at each BS. When $M\geq 2$, Problem \ref{prob:opt_coding} is a challenging  discrete optimization problem with a complex objective function. The number of possible choices for $\mathbf s$ is given by $O\left((M+1)^N\right)$. Thus, a brute-force solution to  Problem \ref{prob:opt_coding}, i.e., exhaustive search, is not acceptable when  $N$ and $M$ are large. In the following, we consider $M\geq 2$.
We aim to obtain a low-complexity solution with superior performance, by carefully  exploiting  structural properties of Problem \ref{prob:opt_coding}.
First, we convert Problem~\ref{prob:opt_coding} into an MCKP, which is a generalization of the ordinary knapsack problem. In particular, there are $N$ mutually disjoint classes $\mathcal C_1,\mathcal C_2,\cdots,\mathcal C_N$, each containing $M+1$ items, to be packaged into a knapsack of capacity $K$. 
Item $m\in\mathcal M\triangleq\{1,2,\cdots,M\}$ in class $\mathcal C_n$ represents that the amount of storage allocated to file $n$ is $\frac{1}{m}$, and item $M+1$ in class $\mathcal C_n$ represents that the amount of storage allocated to file $n$ is $0$.
The profit $p_{n,m}$ and weight $w_{n,m}$ of item $m\in\mathcal M_+\triangleq\{1,2,\cdots,M+1\}$ in class $\mathcal C_n$ are given by
\small{\begin{align}
&p_{n,m}=
\begin{cases}
a_nh(\frac{1}{m},m), &\quad m\in\mathcal M\\
0, &\quad m=M+1
\end{cases},\label{eqn:p_n_m}\\
&w_{n,m}=
\begin{cases}
\frac{1}{m}, &\quad m\in\mathcal M\\
0, &\quad m=M+1
\end{cases}.\label{eqn:w_n_m}
\end{align}}\normalsize
Note that $w_{n,m}$, $n\in\mathcal N$ are the same for all $m\in\mathcal M_+$. Let $x_{n,m}\in\{0,1\}$ represent whether item $m$ in class $\mathcal C_n$ is packaged into the knapsack, where $x_{n,m}=1$ indicates that item $m$ in class $\mathcal C_n$ is packaged into the knapsack.
Thus, the profit sum is given by
\small{\begin{align}
\widetilde{q^c}(\mathbf x)\triangleq\sum_{n\in\mathcal N}\sum_{m\in\mathcal M_+}p_{n,m}x_{n,m},\label{eqn:q^c_x}
\end{align}}\normalsize
and the weight sum is given by
\small{\begin{align}
\sum_{n\in\mathcal N}\sum_{m\in\mathcal M_+}w_{n,m}x_{n,m}.\label{eqn:sum_weight}
\end{align}}\normalsize
Therefore, Problem~\ref{prob:opt_coding} can be transformed into the following MCKP, which chooses
exactly one item from each class such that the profit sum in \eqref{eqn:q^c_x}
is maximized without exceeding the capacity $K$ in the corresponding weight sum in \eqref{eqn:sum_weight}.
\begin{Prob} [Equivalent Problem of Problem \ref{prob:opt_coding} (MCKP)]\label{prob:opt_coding_equi}
\small{\begin{align}
q^{c*}=\max_{\mathbf{x}} &\quad  \widetilde{q^c}(\mathbf x)\nonumber\\
s.t. & \quad \sum_{n\in\mathcal N}\sum_{m\in\mathcal M_+}w_{n,m}x_{n,m}\leq K,\notag\\
    & \quad \sum_{m\in\mathcal M_+}x_{n,m}=1, \quad n\in\mathcal N,\notag\\
    & \quad x_{n,m}\in\{0,1\}, \quad n\in\mathcal N,\ m\in\mathcal M_+,\notag
\end{align}}\normalsize
where $\widetilde{q^c}(\mathbf x)$ is given by \eqref{eqn:q^c_x}. Let $\mathbf x^*$ denote the optimal solution.
\end{Prob}


MCKP is an NP-hard problem, which can be easily shown by reduction from the ordinary knapsack problem. Like other Knapsack Problem variants, MCKP can be solved optimally using two approaches, i.e., the branch-bound method and dynamic programming, with non-polynomial complexity \cite{MCKP10Zhong}. The increase in the number of items will cause the optimization complexity to increase rapidly.
Thus, approximate solutions of polynomial complexity are widely adopted.
For instance, the dynamic programming-based approximate solution proposed in \cite{MCKP04Bansal} can achieve a performance that is no less than $1-\epsilon$ times the optimal value with running time polynomial to $\frac{1}{\epsilon}$, where $\epsilon \in(0,1)$.
Due to the integer requirement (i.e., both the weights and the profits must be positive integers), the dynamic programming-based approximate solution in \cite{MCKP04Bansal} cannot be applied to  Problem~\ref{prob:opt_coding_equi}.
In addition, the greedy solution proposed in \cite{book04Hans} can achieve a performance that is no less than $\frac{1}{2}$ times the optimal value with complexity $O((M+1)N{\rm log}((M+1)N))$.
In the following, we adopt the greedy method in \cite{book04Hans} to obtain a near optimal solution to Problem~\ref{prob:opt_coding_equi} with $\frac{1}{2}$ approximation guarantee.



Before adopting the greedy solution in \cite{book04Hans}, we first introduce some key definitions.
\begin{Def}
\cite{book04Hans} If two items $i$ and $j$ in the same class $\mathcal C_n$ satisfy $w_{n,i}\leq w_{n,j}\ \text{and}\ p_{n,i}\geq p_{n,j}$,
then item $j$ is dominated by item $i$. If three items $i,j,k$ in the same class $\mathcal C_n$ with $w_{n,i}<w_{n,j}<w_{n,k}$ and $p_{n,i}<p_{n,j}<p_{n,k}$ satisfy $\frac{p_{n,k}-p_{n,j}}{w_{n,k}-w_{n,j}}\geq \frac{p_{n,j}-p_{n,i}}{w_{n,j}-w_{n,i}}$,
then item $j$ is LP-dominated by items $i$ and $k$.
\end{Def}


By \eqref{eqn:p_n_m} and \eqref{eqn:w_n_m}, we know that the indices of the dominated and LP-dominated items in each class are the same, and item $M+1$ in each class is not dominated or LP-dominated by any item in the same class. Let $\mathcal R$ denote the set of the indices of undominated items, and
denote $m^+\triangleq \min\{k|k\in\mathcal R, k>m\}$ for all $m\in\mathcal R\setminus\{M+1\}$. In addition, by \cite{book04Hans}, we know that if item $m$ in class $\mathcal C_n$ is dominated by any item in the same class, then an optimal solution to MCKP with $x_{n,m}=0$ exists; if item $m$ in class $\mathcal C_n$ is LP-dominated by any two items in the same class, then an optimal solution to the linear relaxation of MCKP with $x_{n,m}=0$ exists.
Based on these optimality properties, a greedy method is proposed in \cite{book04Hans} to solve MCKP. We adopt this greedy method to solve Problem~\ref{prob:opt_coding_equi}, as summarized in Algorithm~\ref{alg:two_stage}. The complexity of Algorithm~\ref{alg:two_stage} is $O((M+1)N{\rm log}((M+1)N))$. The feasible solution $\mathbf x^{\dagger}$ to Problem~\ref{prob:opt_coding_equi} obtained by Algorithm~\ref{alg:two_stage} achieves a performance that is no less than $\frac{1}{2}$ times the optimal value to Problem~\ref{prob:opt_coding_equi}, i.e., $\widetilde {q^{c}}(\mathbf x^{\dagger})\geq \frac{q^{c*}}{2}$.
\begin{algorithm} \caption{Near Optimal Solution to Problem~\ref{prob:opt_coding_equi} (Problem~\ref{prob:opt_coding})}
\small{\begin{algorithmic}[1]
\STATE Find the set of the indices of undominated items $\mathcal R$.
\STATE Set $x_{n,M+1}=1$, $x_{n,m}=0$ for all $n\in\mathcal N$, $m\in\mathcal M$,  and set the weight sum $W=\sum_{n\in\mathcal N}w_{n,M+1}$ and the profit sum $P=\sum_{n\in\mathcal N}p_{n,M+1}$.
\STATE For all $n\in\mathcal N$ and $m\in\mathcal R\setminus\{M+1\}$, define slope $\lambda_{n,m}=\frac{p_{n,m}-p_{n,m^+}}{w_{n,m}-w_{n,m^+}}$. Order the slopes in $\{\lambda_{n,m}|n\in\mathcal N, m\in\mathcal R\setminus\{M+1\}\}$ in nondecreasing order. Let $\lambda(l)$ denote the $l$-th largest slope.
\STATE Set $l=1$. Let $n,\ m$ be the indices satisfying $\lambda_{n,m}=\lambda(l)$.
\WHILE{$W+w_{n,m}\leq K$}
\STATE Set $x_{n,m}=1$, $x_{n,m^+}=0$, and update $W=W+w_{n,m}-w_{n,m^+}$ and $P=P+p_{n,m}-p_{n,m^+}$.
\STATE Update $l=l+1$. Let $n,\ m$ be the indices satisfying $\lambda_{n,m}=\lambda(l)$.
\ENDWHILE
\IF {W=K}
\STATE Set $\mathbf x^{\dagger}=\mathbf x$.
\ELSE
\STATE
Construct a feasible solution $\overline{\mathbf x}\triangleq(\overline{x}_{i,j})_{i\in\mathcal N,j\in\mathcal M_+}$ to Problem~\ref{prob:opt_coding_equi} by setting $\overline{x}_{n,m}=1$, $\overline{x}_{i,j}=0$ for $i\in\mathcal N$, $i\neq n$ or $j\in\mathcal M_+$, $j\neq m$.
\STATE Set $\mathbf x^{\dagger}= \arg\max_{\mathbf y\in\{\mathbf x, \overline{\mathbf x}\}}\widetilde{q^c}(\mathbf y)$.
\ENDIF
\end{algorithmic}}\normalsize\label{alg:two_stage}
\end{algorithm}

Note that Step $1$ can be conducted using the prune and search method proposed in \cite{David86SIAM}.
In Step $2$, as an initialization, we choose the lightest item $M+1$ for each class.
In Step $3$, the slope $\lambda_{n,m}=\frac{p_{n,m}-p_{n,m^+}}{w_{n,m}-w_{n,m^+}}$, where $m\in\mathcal R\setminus\{M+1\}$, is a measure of the profit-to-weight ratio obtained by choosing item $m$ instead of item $m^+$ in class $\mathcal C_n$.
In Steps $4-8$, items are chosen in the greedy manner according to their slopes.
In Steps $9-14$, we construct a near optimal solution $\mathbf x^{\dagger}$ to Problem~\ref{prob:opt_coding_equi} with $\frac{1}{2}$ approximation guarantee.
In particular, if $W=K$, $\mathbf x^{\dagger}$ is an optimal solution to Problem~\ref{prob:opt_coding_equi}. Otherwise, $\mathbf x^{\dagger}$ is a feasible solution to Problem~\ref{prob:opt_coding_equi} with worst-case performance $\frac{1}{2}$.


\subsubsection{Performance Optimization in Small File Size Regime}

In this part,  we consider the optimization of the asymptotic successful transmission probability $q_0^c(\mathbf s)$ in the small file size regime.
\begin{Prob} [RLNC Caching Design in Small File Size Regime]\label{prob:opt_coding_asymp_0}
\small{\begin{align}
q_0^{c*}\triangleq \max_{\mathbf{s}} &\quad  q_0^c(\mathbf s)\notag\\
s.t. &\quad \eqref{eqn:cache_size_constr_coding}, \eqref{eqn:division_choise_req},\notag
\end{align}}\normalsize
where $q_0^c(\mathbf s)$ is given by \eqref{eqn:CP_coding_caching_aympt_0}.
\end{Prob}

By exploring structural properties of $q_0^c(\mathbf s)$, we can obtain the optimal caching design in the small file size regime.
\begin{Lem}[Optimal Solution to Problem \ref{prob:opt_coding_asymp_0}]\label{Lem:opt_asymp_0}
Suppose $KM\leq N$. There exists $S_0>0$, such that for all $S<S_0$,
the optimal solution to Problem \ref{prob:opt_coding_asymp_0} is given by
\small{\begin{align}
s_n^*=
\begin{cases}
\frac{1}{M}, \quad&n\leq KM\\
0, \quad& n>KM
\end{cases},
\quad n\in\mathcal N,\label{eqn:opt_s_asymp_0}
\end{align}}\normalsize
and the optimal value to Problem \ref{prob:opt_coding_asymp_0} is given by
\small{\begin{align}
q_0^{c*} = \left(1-\frac{{\rm ln}(2)(1+M)S}{(\alpha-2)WT}\right)\sum_{n\in\{1,2,\cdots,KM\}}a_n\;.\label{eqn:STP_saymp_0}
\end{align}}\normalsize
\end{Lem}
\begin{IEEEproof}
Please refer to Appendix D.
\end{IEEEproof}

Lemma \ref{Lem:opt_asymp_0} indicates  that in the small file size regime, when $KM\leq N$,
it is optimal to allocate the storage of each BS equally to the most $KM$ popular files. That is, each of the most $KM$ popular files is partitioned into $M$ subfiles (each of $\frac{S}{M}$ bits), and each BS stores a coded subfile of each of the most $KM$ popular files.
The reason is as follows. In the small file size regime, the  probability that $u_0$ can decode the signal from each of the $M$ nearest BSs is high, and allocating the storage of each of the $M$ nearest BSs equally to $KM$ files maximizes the number of files that can be successfully decoded by $u_0$. Thus, storing  the most $KM$ popular files obviously maximizes the successful transmission probability.
In addition, Lemma \ref{Lem:opt_asymp_0} reveals that in the small file size regime, the optimal successful transmission probability increases with the product of the cache size and SIC capability, i.e., $KM$.

\subsubsection{Performance Optimization in Large File Size Regime}

In this part,  we consider the optimization of the asymptotic successful transmission probability $q_{\infty}^c(\mathbf s)$ in the large file size regime.
\begin{Prob} [RLNC Caching Design in Large  File Size Regime]\label{prob:opt_coding_asymp_infty}
\small{\begin{align}
q_{\infty}^{c*}\triangleq \max_{\mathbf{s}} &\quad  q_{\infty}^c(\mathbf s)\notag\\
s.t. &\quad \eqref{eqn:cache_size_constr_coding}, \eqref{eqn:division_choise_req},\notag
\end{align}}\normalsize
where $q_{\infty}^c(\mathbf s)$ is given by \eqref{eqn:CP_coding_caching_aympt}.
\end{Prob}

By exploring structural properties of $q_{\infty}^c(\mathbf s)$, we can obtain the optimal caching design in the large file size regime.
\begin{Lem}[Optimal Solution to Problem \ref{prob:opt_coding_asymp_infty}]\label{Lem:opt_asymp_infty}
There exists $S_{\infty}>0$, such that for all $S>S_{\infty}$, the optimal solution to Problem \ref{prob:opt_coding_asymp_infty} is given by
\small{\begin{align}
s_n^*=
\begin{cases}
1, \quad&n\leq K\\
0, \quad& n>K
\end{cases},
\quad n\in\mathcal N,\label{eqn:opt_s_asymp_inf}
\end{align}}\normalsize
and the optimal value to Problem \ref{prob:opt_coding_asymp_infty} is given by
\small{\begin{align}
q_{\infty}^{c*} = \frac{2^{-\frac{2S}{\alpha WT}}}{\frac{2}{\alpha}B\left(\frac{2}{\alpha},1-\frac{2}{\alpha}\right)}\sum_{n\in\{1,2,\cdots,K\}}a_n\;.\label{eqn:STP_saymp_infty}
\end{align}}\normalsize
\end{Lem}
\begin{IEEEproof}
Please refer to Appendix E.
\end{IEEEproof}

Lemma \ref{Lem:opt_asymp_infty} indicates that in the large file size regime,
it is optimal to allocate the storage of each BS equally to the most $K$ popular files.
That is, each BS stores each of the most $K$ popular (entire) files.
The reason is as follows. In the large file size regime, the  probability that $u_0$ can decode the signal from any BS besides the nearest one is very small.
Allocating the storage of the nearest BS  to $K$ uncoded files maximizes the number of files that can be successfully decoded by $u_0$. Storing the most $K$ popular files obviously maximizes the successful transmission probability.
In addition, Lemma~\ref{Lem:opt_asymp_infty} reveals that in the large file size regime, the optimal successful transmission probability increases with cache size $K$ and is not affected by SIC capability $M$.

\begin{figure}[t]
\begin{center}
\includegraphics[width=6cm]{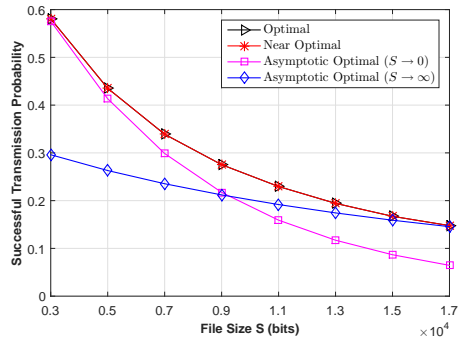}
\end{center}
\caption{Successful transmission probability under the RLNC caching design versus file size $S$. $N=1000$, $K=200$, $\alpha=4$, $W = 10$MHz, $T = 1$ms, $M=4$, and  $a_n=\frac{n^{-\gamma}}{\sum_{n\in \mathcal N}n^{-\gamma}}$ with $\gamma=1$.}
\label{fig:compare_asymp_opt_and_opt}
\end{figure}

Now, we use a numerical example to compare the optimal solution  obtained by  exhaustive search   and the proposed near optimal solution obtained by Algorithm \ref{alg:two_stage} in both successful transmission probability and computational complexity.
We also use this example to verify the asymptotically optimal solutions obtained in Lemmas \ref{Lem:opt_asymp_0} and \ref{Lem:opt_asymp_infty} in the asymptotic file size regimes.
Fig. \ref{fig:compare_asymp_opt_and_opt} plots the successful transmission probability versus file size $S$.
We can see that the successful transmission probability of the proposed near optimal solution  is very close to that of the optimal solution. While, the average computation time for the optimal solution is $3136$ times of that for the near optimal solution. This demonstrates the applicability and effectiveness of the near optimal solution. In addition, we can see that the successful transmission probabilities of the asymptotically optimal solutions obtained by Lemmas \ref{Lem:opt_asymp_0} and \ref{Lem:opt_asymp_infty} approach those of the optimal solutions in the small and large file size regimes, respectively, verifying Lemmas \ref{Lem:opt_asymp_0} and \ref{Lem:opt_asymp_infty}.

\section{Performance Analysis and Optimization of UC caching design}

In this section, we consider the performance analysis and optimization of the UC caching design. First, we analyze the successful transmission probabilities in the general file size regime, the small file size regime and the large file size regime, respectively. Then, we optimize the successful transmission probabilities in these regions.

\subsection{Performance Analysis of UC Caching Design}



\subsubsection{Performance Analysis in General File Size Regime}

According to the total probability theorem, from \eqref{eqn:q_n^u_orig}, we have
\small{\begin{align}
q_n^u(s_n,m_n)=\sum\limits_{i=\frac{1}{s_n}}^{m_n}h(s_n,i)\Pr\left[I_n^u=i\right],\quad &(s_n,m_n)\in\mathcal U\setminus\{(0,0)\},\label{eqn:q_n^u}
\end{align}}\normalsize
where $h(s_n,i)$ is given by  \eqref{eqn:h_z_i}.
To calculate $q_n^u(s_n,m_n)$, it remains to calculate the p.m.f. of $I_n^u$.
Collecting $\frac{1}{s_n}$ different subfiles of file $n$ under the UC caching design can be viewed as a classical Coupon Collector's Problem, where $\frac{1}{s_n}$ distinct objects (i.e., coupons) are repeatedly drawn (with replacement) from an urn with probability $s_n$ of picking an object at each trial until each of the $\frac{1}{s_n}$ objects is picked at least once.
In particular, $\frac{1}{s_n}$ different subfiles of file $n$ can be regarded as $\frac{1}{s_n}$ different objects in the Coupon Collector's Problem. The subfile of file $n$ stored at BS $i\in\{1,2,\cdots,m_n\}$ can be regarded as the object drawn at the $i$-th trial in the Coupon Collector's Problem.
Recovering file $n$ (i.e., obtaining $\frac{1}{s_n}$ different subfiles of file $n$) from the subfiles of file $n$ stored at the $i$ nearest BSs is equivalent to collecting  all $\frac{1}{s_n}$ types of objects within $i$ trials in the Coupon Collector's Problem.
Thus, $I_n^u$ can be regarded as the minimum number of trials needed to get all $\frac{1}{s_n}$ distinct objects in the Coupon Collector's Problem. From the results of the Coupon Collector's Problem \cite{CCP_web}, we have
\small{\begin{align}
&p\left(s_n,i\right)\triangleq\mathbb{P} [I_n^u=i]
=\begin{cases}
1, &\quad s_n=1,\;i=1\\
0, &\quad s_n=1,\;i\in\{2,3,\cdots,M\}\\
\sum\limits_{k=0}^{\frac{1}{s_n}-1}(-1)^k\binom{\frac{1}{s_n}-1}{k}(1-s_n(k+1))^{i-1}, &\quad s_n\in\mathcal S\setminus\{0,1\},\; i\in\{\frac{1}{s_n},\frac{1}{s_n}+1,\cdots,M\}
\end{cases}.\label{eqn:p_m_n_i_uncoded_caching}
\end{align}}\normalsize
From \eqref{eqn:q_n^u_orig}, \eqref{eqn:h_z_i}, \eqref{eqn:q_n^u} and \eqref{eqn:p_m_n_i_uncoded_caching}, we have
\small{\begin{align}
q_n^u(s_n,m_n)=
\begin{cases}
0,\quad &(s_n,m_n)=(0,0)\\
\sum_{i=\frac{1}{s_n}}^{m_n}p(s_n,i)h(s_n,i), \quad &(s_n,m_n)\in\mathcal U\setminus(0,0)
\end{cases}.\label{eqn:q_n^u_final}
\end{align}}\normalsize
Therefore, we have the successful transmission probability under the UC caching design, as summarized below.
\begin{Thm}[Performance of UC in General File Size Regime]\label{Thm:CP_close_form_partition}
The successful transmission probability $q^u(\mathbf s,\mathbf m)$ under the UC caching design is given by
\small{\begin{align}
q^u(\mathbf s,\mathbf m)&=\sum_{n\in\mathcal N}a_nq_n^u(s_n,m_n),\quad (\mathbf s,\mathbf m)\in\mathcal U^{N},\label{eqn:CP_partition_caching_close_form}
\end{align}}\normalsize
where $q_n^u(s_n,m_n)$ is given by \eqref{eqn:q_n^u_final}.
\end{Thm}

\begin{figure}[t]
\begin{center}
\includegraphics[width=6cm]{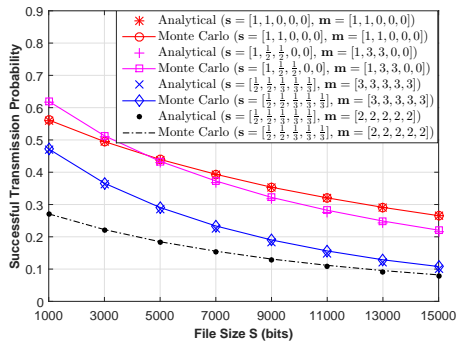}
\end{center}
\caption{Successful transmission probability under the UC caching design versus file size $S$. $N=5$, $K=2$, $\alpha=4$, $W = 10$MHz, $T = 1$ms, $M = 3$, and $a_n=\frac{n^{-\gamma}}{\sum_{n\in \mathcal N}n^{-\gamma}}$ with $\gamma=1$.}
\label{fig:verification_anal_MC_partition}
\end{figure}

From Theorem \ref{Thm:CP_close_form_partition}, we can see that $q^u(\mathbf s,\mathbf m)$ is a decreasing function of $S$ and an increasing function of $m_n$ for all $n\in\mathcal N$.
The impact  of $\mathbf s$ on $q^u(\mathbf s,\mathbf m)$ is not obvious.
Fig.~\ref{fig:verification_anal_MC_partition} plots $q^u(\mathbf s,\mathbf m)$ versus $S$ at different $\mathbf s$. Fig. \ref{fig:verification_anal_MC_partition} verifies Theorem \ref{Thm:CP_close_form_partition} and demonstrates the accuracy of the approximation adopted.
In addition, from Fig. \ref{fig:verification_anal_MC_partition}, we can see that $q^u(\mathbf s,\mathbf m)$ decreases with $S$ and increases with $m_n$ for all $n\in\mathcal N$. The impact of $\mathbf s$ on $q^u(\mathbf s,\mathbf m)$ is not clear in the general file size regime.

By comparing Theorem~\ref{Thm:CP_close_form_dist_order} and Theorem~\ref{Thm:CP_close_form_partition}, we have the following corollary.
\begin{Cor}[Performance Comparison between RLNC and UC Caching Designs]\label{cor:comparison}
Given any caching design parameter $\mathbf s\in\mathbf S^N$, the successful transmission probability under the RLNC caching design is greater than or equal to that under the UC caching design, i.e., $q^c(\mathbf s)\geq q^u(\mathbf s,\mathbf m)$ for all $\mathbf m$ such that $(\mathbf s, \mathbf m)\in\mathcal U^N$, where the equality holds when $s_n\in\{0,1\}$ for all $n\in\mathcal N$.
\end{Cor}

\subsubsection{Performance Analysis in Small File Size Regime}

From  Theorem \ref{Thm:CP_close_form_partition}, we derive the asymptotic successful transmission probability in the small file size regime (i.e., $S\to 0$) as follows.
\begin{Lem}[Performance of UC in Small File Size Regime]\label{Lem:CP_partition_asymp_0}
For all $(\mathbf s,\mathbf m)\in\mathcal U^{N}$, we have $q^u(\mathbf s,\mathbf m)\stackrel{S\to0}{\sim}q_0^u(\mathbf s,\mathbf m)$, where
\small{\begin{align}
q_0^u(\mathbf s,\mathbf m)\triangleq &\sum_{n\in\mathcal N}a_n\sum_{i=\frac{1}{s_n}}^{m_n}p\left(s_n,i\right)
-\frac{{\rm ln}(2)}{(\alpha-2)WT}S\sum_{n\in\mathcal N}a_ns_n\sum_{i=\frac{1}{s_n}}^{m_n}p\left(s_n,i\right)(i+1)i.\label{eqn:CP_partition_caching_aympt_0}
\end{align}}\normalsize
Here, $p(\cdot,\cdot)$ is given by \eqref{eqn:p_m_n_i_uncoded_caching}.
\end{Lem}
\begin{IEEEproof}
Lemma~\ref{Lem:CP_partition_asymp_0} can be proved in a similar way to Lemma~\ref{Lem:CP_asymp_0}. We omit the details due to page limitation.
\end{IEEEproof}

From Lemma~\ref{Lem:CP_partition_asymp_0}, we know that $\lim_{S\to 0}q^u(\mathbf s,\mathbf m)=\sum_{n\in\mathcal N}a_n\sum_{i=\frac{1}{s_n}}^{m_n}p\left(s_n,i\right)$, which represents the success probability of collecting all different subfiles of a randomly requested file, and $q_0^u(\mathbf s,\mathbf m)$ increases linearly to $\sum_{n\in\mathcal N}a_n\sum_{i=\frac{1}{s_n}}^{m_n}p\left(s_n,i\right)$ as $S$ decreases to $0$.
In addition, $\mathbf s$  affects the asymptotic behavior of $q_0^u(\mathbf s,\mathbf m)$ by affecting $\lim_{S\to 0}q^u(\mathbf s,\mathbf m)$ and the coefficient of $S$.
Fig.~\ref{fig:asymp_STP_partition}~(a) plots $q_0^u(\mathbf s,\mathbf m)$ versus $S$ in the small file size regime. We see from Fig.~\ref{fig:asymp_STP_partition}~(a) that when $S$ decreases, the gap between each ``General" curve, which is plotted using Theorem~\ref{Thm:CP_close_form_partition}, and the corresponding ``Asymptotic" curve, which is plotted using Lemma~\ref{Lem:CP_partition_asymp_0}, decreases. Thus, Fig.~\ref{fig:asymp_STP_partition}~(a) verifies Lemma~\ref{Lem:CP_partition_asymp_0}.

\subsubsection{Performance Analysis in Large File Size Regime}

From Theorem \ref{Thm:CP_close_form_partition}, we derive the asymptotic successful transmission probability in the large file size regime (i.e., $S\to\infty$) as follows.
\begin{Lem}[Performance of UC in Large File Size Regime]\label{Lem:CP_partition_asymp_CP}
For all $(\mathbf s,\mathbf m)\in\mathcal U^{N}$, we have $q^u(\mathbf s,\mathbf m)\stackrel{S\to\infty}{\sim}q_{\infty}^u(\mathbf s)$, where
\small{\begin{align}
q_{\infty}^u(\mathbf s)&\triangleq \frac{2^{-\frac{\left(s_{\max}+1\right)S}{\alpha s_{\max}WT}}p\left(s_{\max},\frac{1}{s_{\max}}\right)}{\left(\frac{2}{\alpha}B\left(\frac{2}{\alpha},1-\frac{2}{\alpha}\right)\right)^{\frac{s_{\max}+1}{2s_{\max}^2}}}\sum_{n\in\mathcal N}a_n \mathbf 1[s_n=s_{\max}].\label{eqn:CP_partition_caching_aympt}
\end{align}}\normalsize
Here, $p(\cdot,\cdot)$ is given by \eqref{eqn:p_m_n_i_uncoded_caching}.
\end{Lem}
\begin{IEEEproof}
Lemma~\ref{Lem:CP_partition_asymp_CP} can be proved in a similar way to Lemma~\ref{Lem:CP_asymp_CP}. We omit the details due to page limitation.
\end{IEEEproof}

\begin{figure}[t]
\begin{center}
 \subfigure[\small{Small file size regime.}]
 {\resizebox{6cm}{!}{\includegraphics{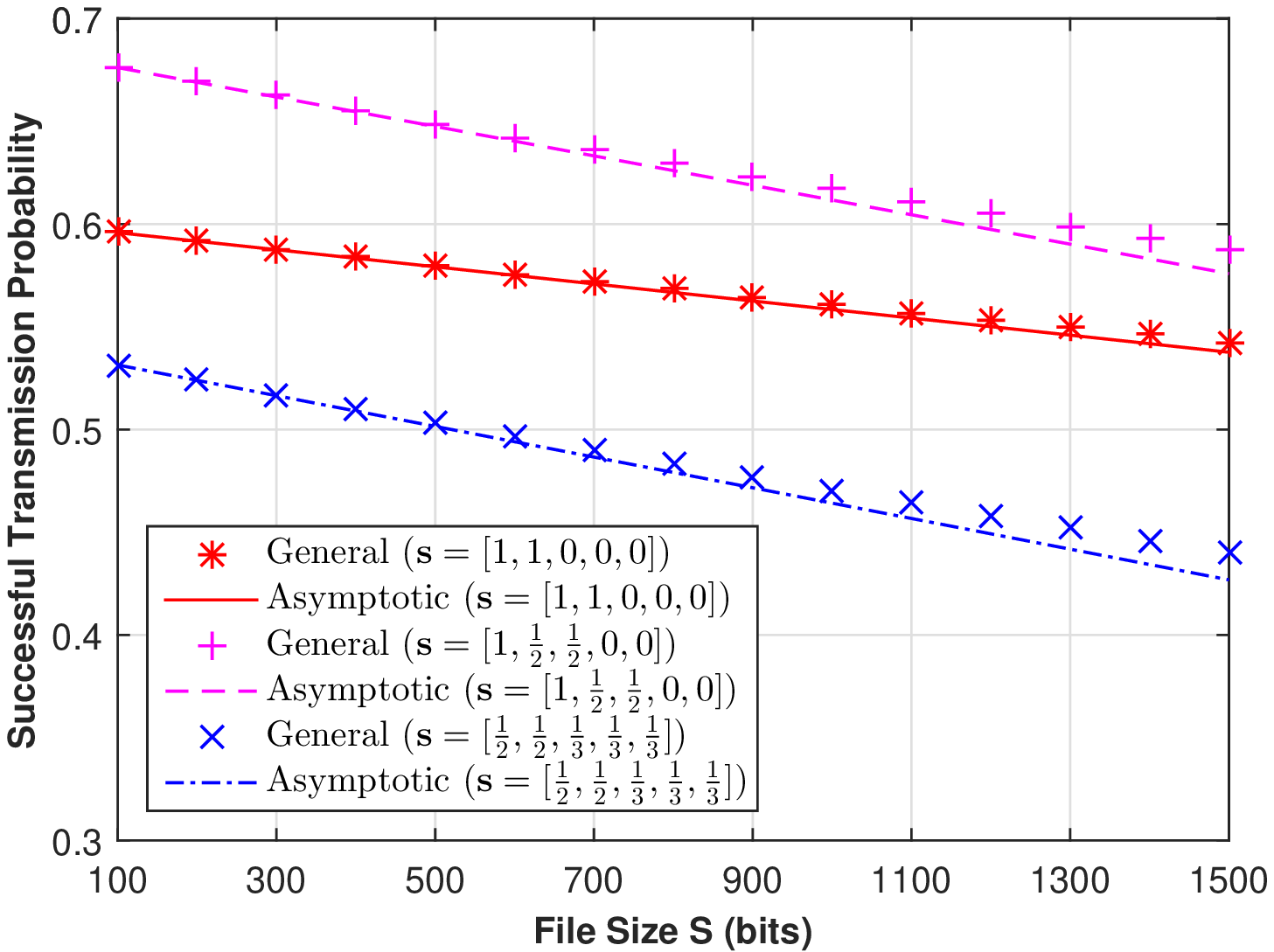}}}\quad\quad
 \subfigure[\small{Large file size regime.}]
 {\resizebox{6cm}{!}{\includegraphics{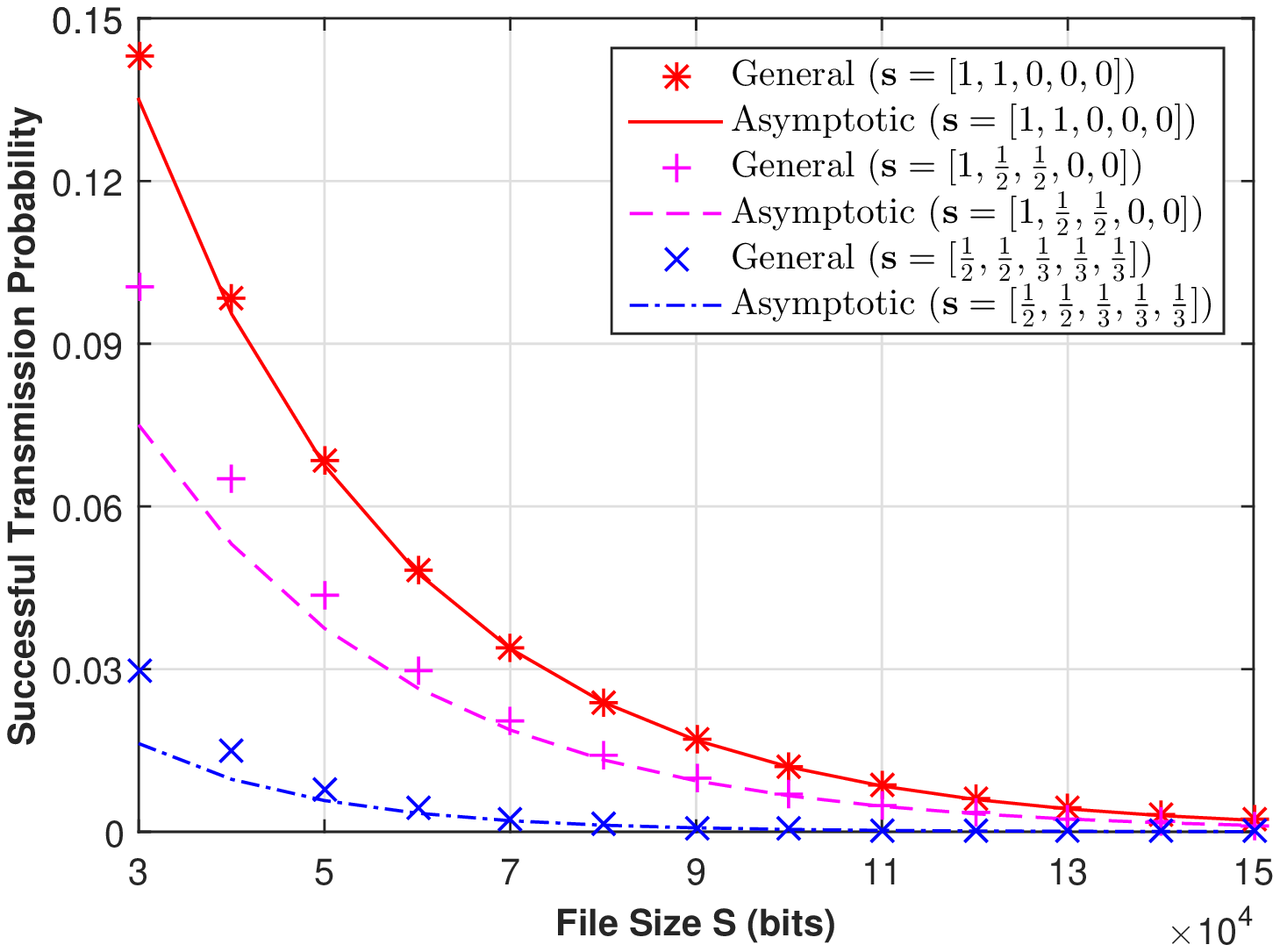}}}
 \end{center}
   \caption{\small{Successful transmission probability under the UC caching design versus file size $S$.  $N=5$, $K=2$, $\alpha=4$, $W = 10$MHz, $T = 1$ms, $M = 3$, and $a_n=\frac{n^{-\gamma}}{\sum_{n\in \mathcal N}n^{-\gamma}}$ with $\gamma=1$.}}
\label{fig:asymp_STP_partition}
\end{figure}

Note that $q_{\infty}^u(\mathbf s)$ does not depend on $\mathbf m$.
From Lemma \ref{Lem:CP_partition_asymp_CP}, we know that $\lim_{S\to \infty}q^u(\mathbf s,\mathbf m)=0$, and $q_{\infty}^u(\mathbf s)$ decreases exponentially to $0$ as $S$ increases to $\infty$.
In addition, $\mathbf s$  affects the asymptotic behavior of $q_{\infty}^u(\mathbf s)$ in the form of $s_{\max}$ only.
By comparing Lemma~\ref{Lem:CP_asymp_CP} and Lemma~\ref{Lem:CP_partition_asymp_CP}, we can see that in the large file size regime, the successful transmission probability under the UC caching design is $p(s_{\max},\frac{1}{s_{\max}})$ of that under the RLNC caching design.
Fig.~\ref{fig:asymp_STP_partition} (b) plots $q_{\infty}^u(\mathbf s)$ versus $S$ in the large file size regime. We see from Fig.~\ref{fig:asymp_STP_partition} (b) that when $S$ increases, the gap between each ``General" curve, which is plotted using Theorem~\ref{Thm:CP_close_form_partition}, and the corresponding ``Asymptotic" curve, which is plotted using Lemma~\ref{Lem:CP_partition_asymp_CP}, decreases. Thus, Fig. \ref{fig:asymp_STP_partition} (b) verifies Lemma~ \ref{Lem:CP_partition_asymp_CP}.

\subsection{Performance Optimization of UC Caching Design}

Note that $q^u(\mathbf s,\mathbf m)$ is an increasing function of $m_n$ for all $n\in\mathcal N$. In the following, to study the optimal successful transmission probability under the UC caching design, we set
\small{\begin{align}
m_n=
\begin{cases}
0, \quad &s_n=0\\
1, &s_n=1\\
M, &\text{otherwise}
\end{cases}.\label{eqn:m_n_set}
\end{align}}\normalsize
Specifically, in the general file size regime, we focus on optimizing $q^u(\mathbf s)\triangleq \sum_{n\in\mathcal N}a_nq_n^u(s_n,m_n)$ over $\mathbf s\in\mathcal S^N$, where $\mathbf m$ is given by \eqref{eqn:m_n_set}.
In the small file size regime, we focus on optimizing $q_0^u(\mathbf s)\triangleq q_0^u(\mathbf s,\mathbf m)$ over $\mathbf s\in\mathcal S^N$, where $\mathbf m$ is given by \eqref{eqn:m_n_set}. In the large file size regime, we focus on optimizing $q_{\infty}^u(\mathbf s)$ over $\mathbf s\in\mathcal S^N$.

\subsubsection{Performance Optimization in General File Size Regime}\label{sec:UC_opt_general}
The caching design affects the successful transmission probability $q^u(\mathbf s)$ via design parameter $\mathbf s$. We would like to maximize $q^u(\mathbf s)$ by carefully optimizing $\mathbf s$.
\begin{Prob} [UC Caching Design in General File Size Regime]\label{prob:opt_coding_partition}
\small{\begin{align}
q^{u*}\triangleq \max_{\mathbf{s}} &\quad  q^u(\mathbf s)\notag\\
s.t. &\quad \eqref{eqn:cache_size_constr_coding}, \eqref{eqn:division_choise_req},\notag
\end{align}}\normalsize
where $q^u(\mathbf s)$ is given by \eqref{eqn:CP_partition_caching_close_form}. Let $\mathbf s^*$ denote the optimal solution.
\end{Prob}

When $M=1$, we have $s_n\in\{0,1\}$ for all $n\in\mathcal N$, and the proposed UC caching design degenerates to deterministic and identical caching of entire files. Based on structural properties of $q^{u}(\mathbf s)$, we know that the optimal caching design is to store the most $K$ popular entire files  at each BS. When $M\geq 2$,
as Problem~\ref{prob:opt_coding}, Problem~\ref{prob:opt_coding_partition} can also be converted into an MCKP and solved with $\frac{1}{2}$ approximation guarantee using the greedy method in \cite{book04Hans}.

\subsubsection{Performance Optimization in Small File Size Regime}

In this part,  we consider the optimization of the asymptotic successful transmission probability $q_0^u(\mathbf s)$ in the small file size regime.
\begin{Prob} [UC Caching Design in Small File Size Regime]\label{prob:opt_partition_asymp_0}
\small{\begin{align}
q_0^{u*}\triangleq \max_{\mathbf{s}} &\quad  q_0^u(\mathbf s)\notag\\
s.t. &\quad \eqref{eqn:cache_size_constr_coding}, \eqref{eqn:division_choise_req},\notag
\end{align}}\normalsize
where $q_0^u(\mathbf s)$ is given by \eqref{eqn:CP_partition_caching_aympt_0}.
\end{Prob}

Similarly, Problem~\ref{prob:opt_partition_asymp_0} can also be converted into an MCKP and solved with $\frac{1}{2}$ approximation guarantee using the greedy method in \cite{book04Hans}. 

\subsubsection{Performance optimization in large file size regime}

In this part,  we consider the optimization of the asymptotic successful transmission probability $q_{\infty}^u(\mathbf s)$ in the large file size regime.
\begin{Prob} [UC Caching Design in Large File Size Regime]\label{prob:opt_partition_asymp_infty}
\small{\begin{align}
q_{\infty}^{u*}\triangleq \max_{\mathbf{s}} &\quad q_{\infty}^u(\mathbf s)\notag\\
s.t. &\quad \eqref{eqn:cache_size_constr_coding}, \eqref{eqn:division_choise_req},\notag
\end{align}}\normalsize
where $q_{\infty}^u(\mathbf s)$ is given by \eqref{eqn:CP_partition_caching_aympt}.
\end{Prob}

By exploring structural properties of $q_{\infty}^u(\mathbf s)$, we can obtain the optimal caching design in the large file size regime.
\begin{Lem}[Optimal Solution to Problem \ref{prob:opt_partition_asymp_infty}]\label{Lem:opt_asymp_infty_partition}
There exists $S_{\infty}>0$, such that for all $S>S_{\infty}$, the optimal solution to Problem \ref{prob:opt_partition_asymp_infty} is given by
\small{\begin{align}
s_n^*=
\begin{cases}
1, \quad&n\leq K\\
0, \quad& n>K
\end{cases},
\quad n\in\mathcal N,\label{eqn:opt_z_asymp_inf}
\end{align}}\normalsize
and the optimal value to Problem \ref{prob:opt_partition_asymp_infty} is given by
\small{\begin{align}
q_{\infty}^{u*} = \frac{2^{-\frac{2S}{\alpha WT}}}{\frac{2}{\alpha}B\left(\frac{2}{\alpha},1-\frac{2}{\alpha}\right)}\sum_{n\in\{1,2,\cdots,K\}}a_n\;.\label{eqn:STP_saymp_infty_partition}
\end{align}}\normalsize
\end{Lem}

Lemma~\ref{Lem:opt_asymp_infty_partition} indicates that in the large file size regime, it is optimal to allocate the storage of each BS equally to the most $K$ popular files. That is, each BS stores each of the most $K$ popular files.
Note that the optimal UC caching design is the same as the optimal RLNC caching design in the large file size regime. This indicates that the advantage of the RLNC caching design over the UC caching design vanishes in the large file size regime.

\section{numerical results}


\begin{figure}[!t]
\begin{center}
 \subfigure[\small{File size at $M=5$, $K=200$, and $\gamma=1$.}]
 {\resizebox{6cm}{!}{\includegraphics{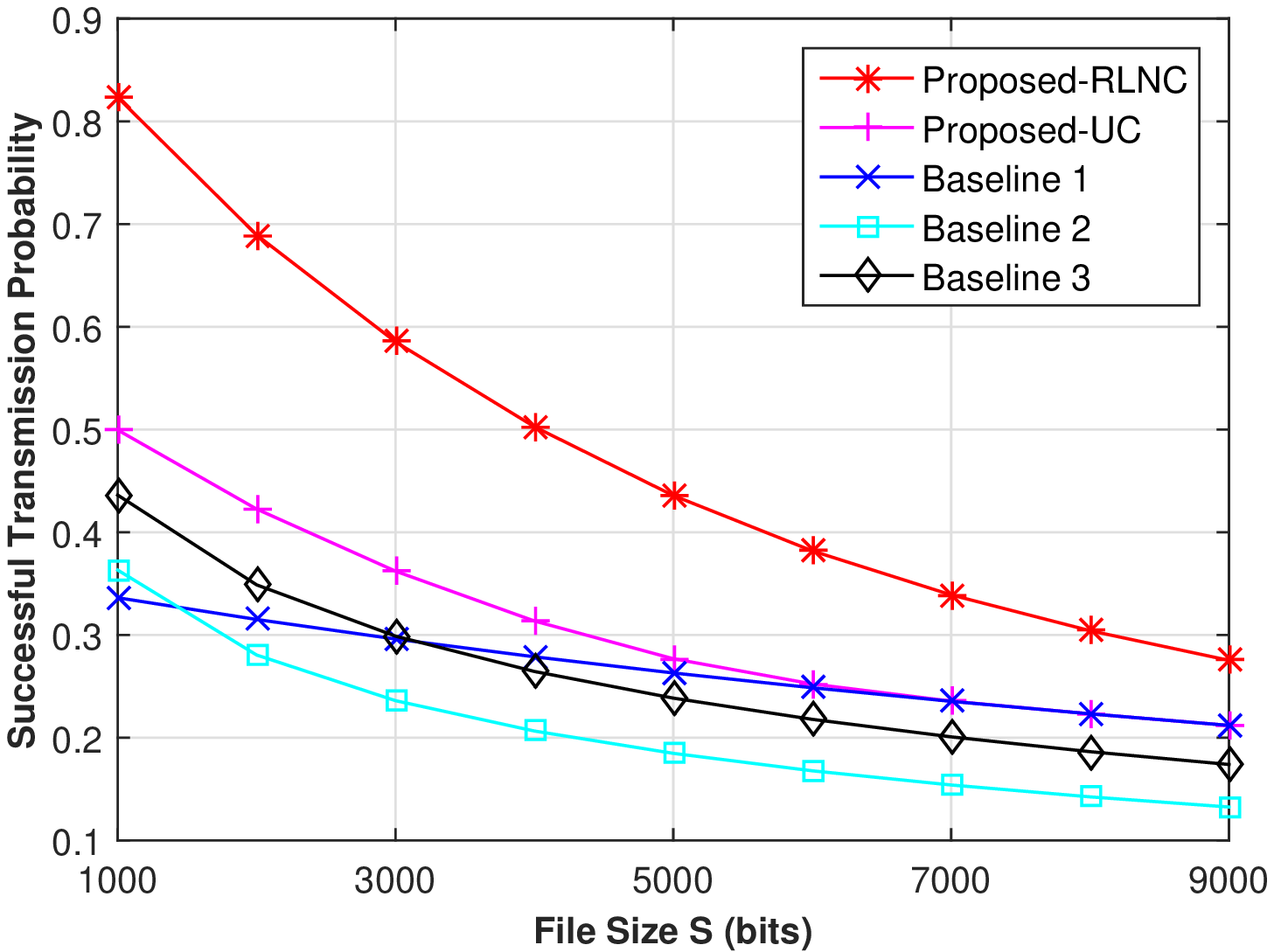}}}\quad\quad
 \subfigure[\small{SIC capability at $S=2\times10^3$, $K=200$, and $\gamma=1$.}]
 {\resizebox{6cm}{!}{\includegraphics{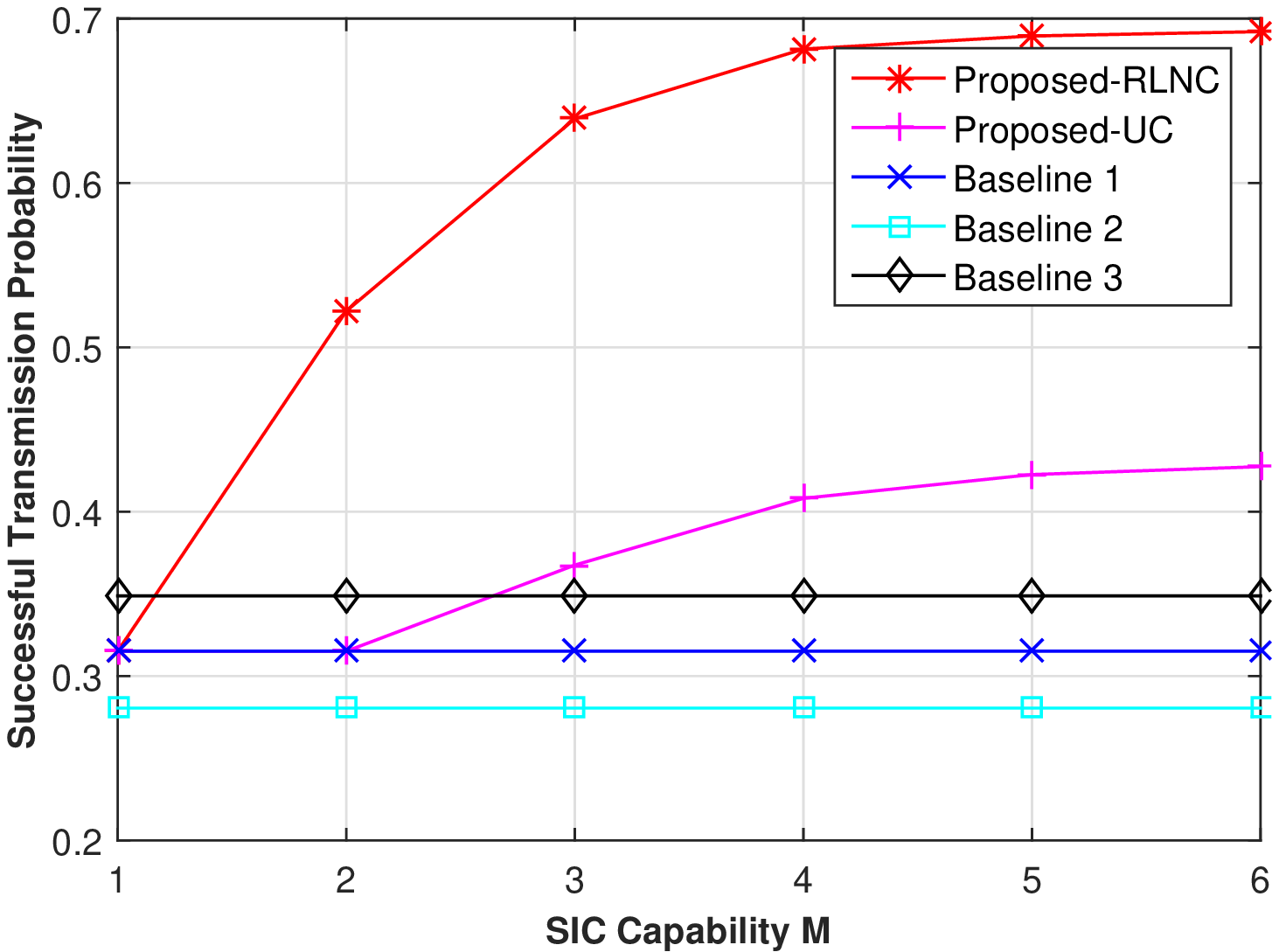}}}
 \subfigure[\small{Cache size at $S=2\times10^3$, $M=5$, and $\gamma=1$.}]
 {\resizebox{6cm}{!}{\includegraphics{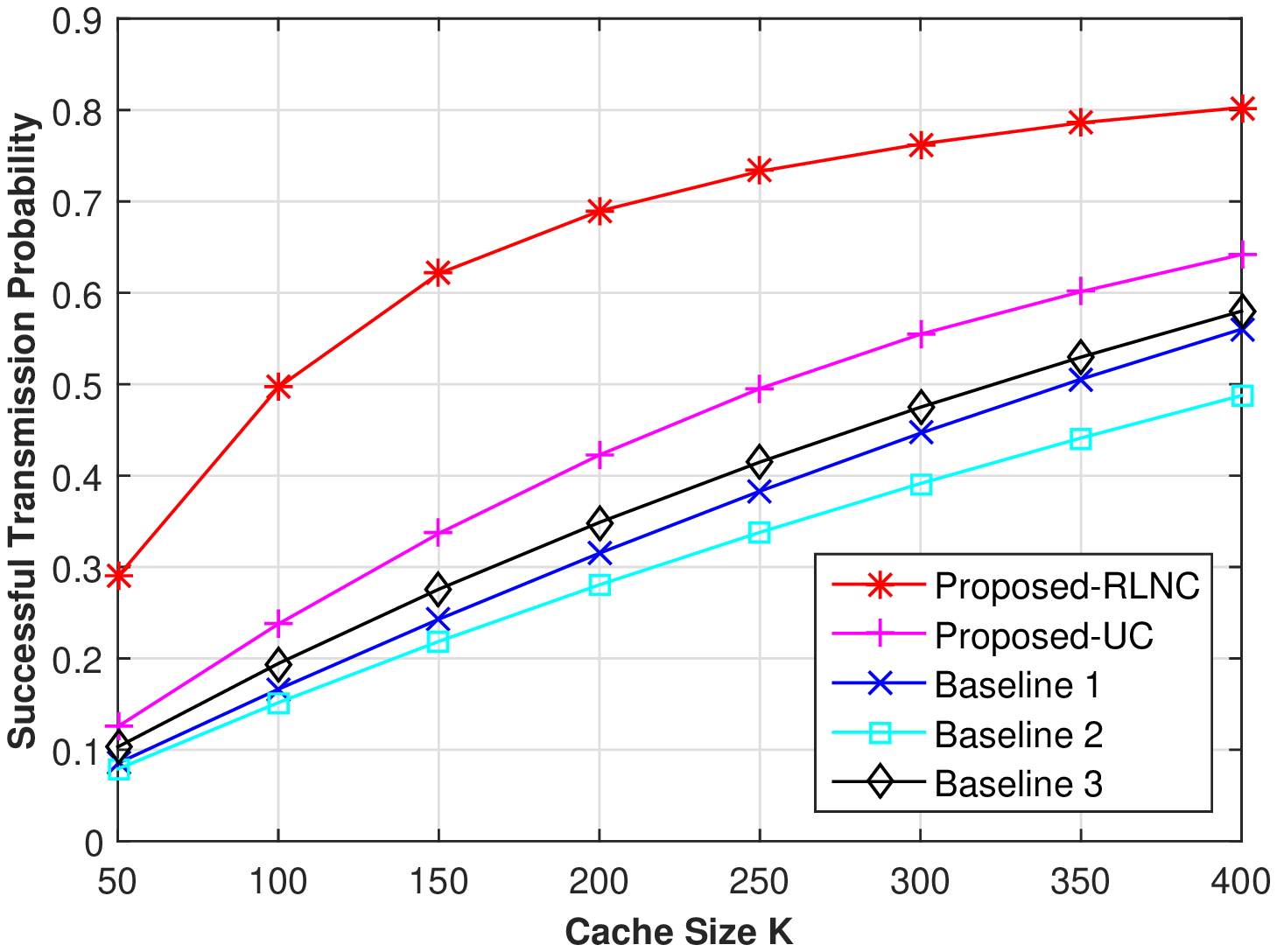}}}\quad\quad
 \subfigure[\small{Zipf exponent at $S=2\times10^3$, $M=5$, and $K=200$.}]
 {\resizebox{6cm}{!}{\includegraphics{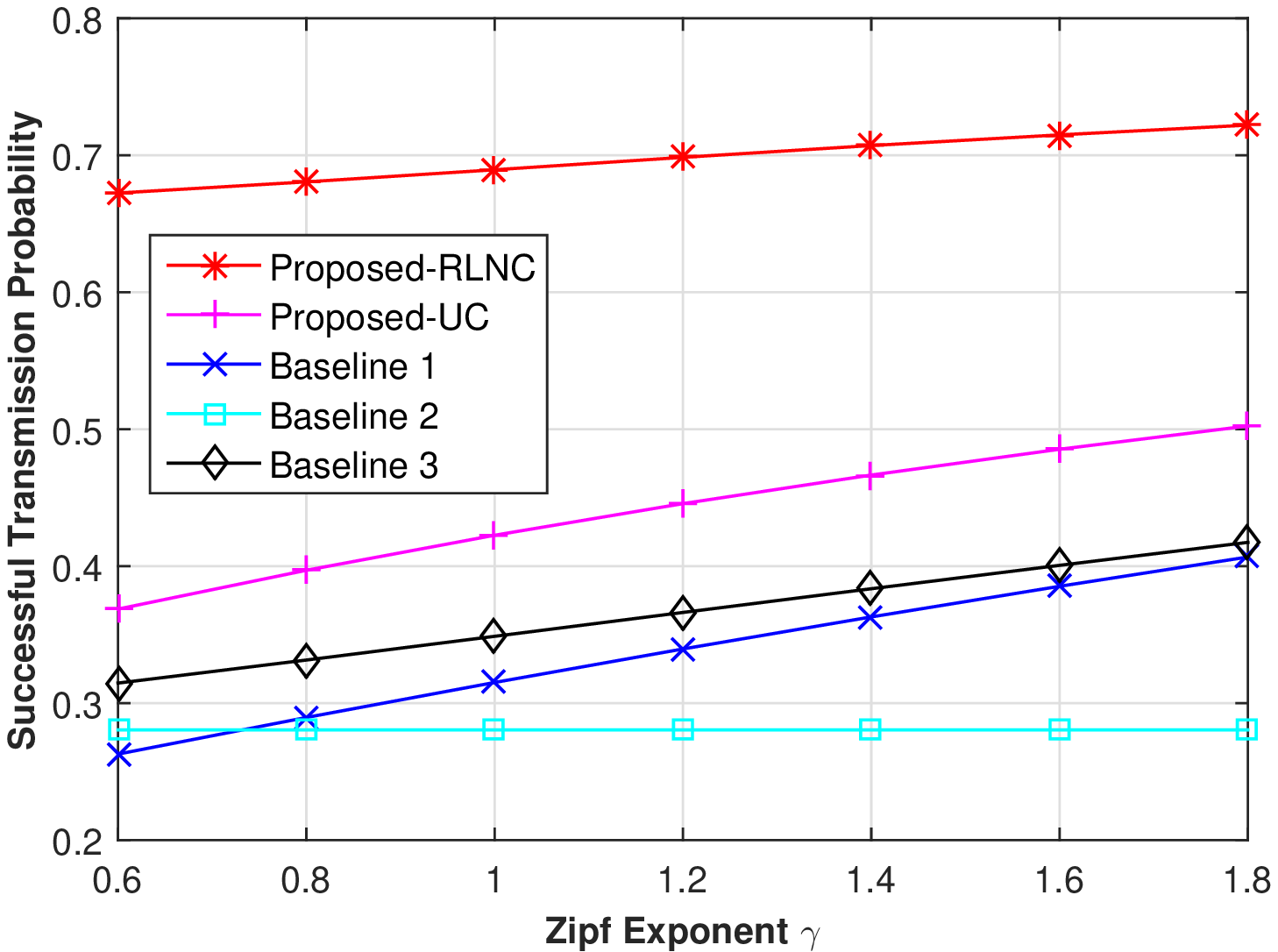}}}
 \end{center}
   \caption{\small{Successful transmission probability versus file size $S$, SIC capability $M$, cache size $K$, and Zipf exponent $\gamma$.}}
\label{fig:simulation}
\end{figure}

In this section, we compare the proposed near optimal RLNC caching design with the proposed near optimal UC caching design and three baselines in the existing literature.
Baseline $1$ refers to the caching design in which the most $K$ popular entire files are stored at each BS (i.e., $s_n=1$ for $n\in\{1,2,\cdots,K\}$, and $s_n=0$ for $n\in\{K+1,K+2,\cdots,N\}$) \cite{EURASIP15Debbah}.
Baseline $2$ refers to the random caching design in which all files in $\mathcal N$ are randomly stored at each BS with equal caching probability $\frac{K}{N}$ \cite{TamoorComLett16}. Baseline $3$ refers to the random caching design in which file $n\in\mathcal N$ is stored at a BS with caching probability $T_n=\min\{a_nK+\mu,1\}$, where $c$ satisfies $\sum_{n\in\mathcal N}\min\{a_nK+\mu,1\}=K$.\footnote{The caching probability $(T_n)_{n\in\mathcal N}$ in Baseline $3$ is the vector which minimizes the distance from the scaled file popularity $K\mathbf a$ under constraints $0\leq T_n\leq 1$ and $\sum_{n\in\mathcal N}T_n=K$.}
Under Baselines $1$, $2$ and $3$, $u_0$  requesting file $n$ is associated with the nearest BS which stores file $n$ \cite{Cui16TWC}.
In the simulation, the popularity follows Zipf distribution, i.e., $a_n=\frac{n^{-\gamma}}{\sum_{n\in\mathcal N}n^{-\gamma}}$, where $\gamma$ is the Zipf exponent.
We choose $N=1000$, $\alpha=4$, $W = 10$MHz and $T = 1$ms.

Fig. \ref{fig:simulation} illustrates the successful transmission probability versus different system parameters. We can observe that the proposed RLNC caching design significantly outperforms the proposed UC caching design and the three baseline designs, and its performance increases much faster with the SIC capability and the cache size. This is because the proposed RLNC caching design wisely exploits SIC capability and storage resource. In addition, the two proposed partition-based caching designs have better performance than the three baseline designs, which focus on storing entire files. This is because the partition-based caching designs achieve higher file diversities than the caching designs storing entire files.

Specifically, Fig.~\ref{fig:simulation} (a) illustrates the successful transmission probability versus the file size $S$. We can see that
the performance gaps between the proposed RLNC caching design and the other caching designs are relative large at small $S$. This is because when $S$ is small, the success probability of sequentially decoding multiple signals using SIC is relatively large, and hence, the benefit of high file diversity offered by the proposed RLNC caching design can be seen more clearly.

Fig.~\ref{fig:simulation} (b) illustrates the successful transmission probability versus the SIC capability $M$. We can see that the performance of the proposed RLNC and UC caching designs increases with $M$,
while the performance of the three baseline designs is not affected by $M$. This is because under the proposed partition-based caching designs, more subfiles can be decoded as $M$ increases; under the three baseline designs, which store entire files, the successful transmission probability of an entire file from the nearest BS does not change with $M$.
In addition, note that when $M=1$, the successful transmission probabilities of the proposed RLNC and UC caching designs and Baseline $1$ are the same. This is because as discussed in Sections~\ref{sec:RLNC_opt_general}  and \ref{sec:UC_opt_general}, the optimal RLNC and UC caching designs degenerate to Baseline $1$, when $M=1$.

Fig.~\ref{fig:simulation} (c) illustrates the successful transmission probability versus the cache size $K$. We can see that the performance of all the caching designs increases with $K$. This is because as $K$ increases, each BS can store more files, and the probability that a randomly requested file can be obtained from the nearby BSs increases.

Fig.~\ref{fig:simulation} (d) illustrates the successful transmission probability versus the Zipf exponent $\gamma$. We can see that the performance of the proposed RLNC and UC caching designs, Baseline $1$ and Baseline $3$ increases with $\gamma$. This is because when $\gamma$ increases, the probability that a requested file is a popular one increases. These popularity-aware caching designs store more popular files in the network, while Baseline $2$  allocates the same storage resource to all the files.

\section{conclusion}

In this paper, we proposed two partition-based caching designs, i.e., a RLNC caching design and an UC caching design, and analyzed and optimized the two caching designs in a large-scale SIC-enabled wireless network.
First, by utilizing tools from stochastic geometry and adopting appropriate approximations, we analyzed the successful transmission probabilities in the general file size regime and the two asymptotic file size regimes.
We also showed that the RLNC caching design outperforms the UC caching design in the general file size regime, and the successful transmission probability of each proposed caching design decreases linearly with the file size in the small file size regime, and decreases exponentially with the file size in the large file size regime.
Then, for each proposed caching design, by exploring structural properties, we successfully transformed the original caching optimization problem, which is NP-hard, into an MCKP problem, and obtained a near optimal solution with $\frac{1}{2}$ approximation guarantee and polynomial complexity in the general file size regime. We also obtained closed-form asymptotically optimal solutions.
The asymptotic optimization results show that in the small file size regime, the optimal successful transmission probability under the RLNC caching design increases with the cache size and the SIC capability. While, in the large file size regime, the optimal successful transmission probability of each proposed caching design increases with the cache size and is not affected by the SIC capability.

\section*{Appendix A: Proof of Lemma~\ref{Lem:exp_of_h_s_i}}\label{sec:proof_A}

First, the conditional probability of successfully decoding the subfile of file $n$ stored at BS $j$ after successfully decoding and canceling the signals from the nearer BSs in $\{1,2,\cdots j-1\}$, conditioned on $d_j=x$, is given by
\small{\begin{align}
p_{n,j,d_j}(s_n,x)&\triangleq \mathbb P\left[W{\rm log}_2(1+{\rm SIR}_j)>\frac{s_nS}{T}|d_j=x\right]=\mathbb P\left[{\rm SIR}_j>2^{\frac{s_nS}{WT}}-1|d_j=x\right]\notag\\
&\eqla \mathbb E_{I_j}\left[\mathbb P\left[|h_j|^2>\left(2^{\frac{s_nS}{WT}}-1\right)d_j^{\alpha}I_j|d_j=x\right]\right]\eqlb \underbrace{\mathbb E_{I_j}\left[\exp(-\left(2^{\frac{s_nS}{WT}}-1\right) x^{\alpha}I_j)\right]}_{\triangleq \mathcal L_{I_j}(s,x)\big{|}_{s=\big(2^{\frac{s_nS}{WT}}-1\big) x^{\alpha}}},\label{eqn:p_m_cond_d_m}
\end{align}}\normalsize
where $(a)$ is obtained based on \eqref{eqn:SIR_m_expression}, and $(b)$ is obtained by noting that $|h_j|^{2}\dis \exp(1)$. To calculate $p_{n,j,d_j}(s_n,x)$ using \eqref{eqn:p_m_cond_d_m}, we only need to calculate $\mathcal L_{I_j}(s,x)$. The expression of $\mathcal L_{I_j}(s,x)$ is calculated as follows
\small{\begin{align}
\mathcal L_{I_j}(s,x)&=\mathbb E\left[\exp\left(-s\sum_{k\in\Phi_b\setminus\{1,2,\cdots,j\}}d_k^{-\alpha}|h_k|^2\right)\right]
=\mathbb E\left[\prod_{k\in\Phi_b\setminus\{1,2,\cdots,j\}}\exp\left(-sd_k^{-\alpha}|h_k|^2\right)\right]\notag
\end{align}
\begin{align}
&\eqlc\exp\left(-2\pi\lambda_b\int_x^{\infty}\left(1-\frac{1}{1+sr^{-\alpha}}\right)r{\rm d}r\right)\eqld \exp\left(-\frac{2\pi}{\alpha}\lambda_bs^{\frac{2}{\alpha}}B'\left(\frac{2}{\alpha},1-\frac{2}{\alpha},\frac{1}{1+sx^{-\alpha}}\right)\right),\label{eqn:laplace_transform_inf_m}
\end{align}}\normalsize
where $(c)$ is obtained by utilizing the probability generating functional of PPP \cite[Page 235]{FTNhaenggi09}, and $(d)$ is obtained by first replacing $s^{-\frac{1}{\alpha}}r$ with $t$, and then replacing $\frac{1}{1+t^{-\alpha}}$ with $w$. Substituting $s=\big(2^{\frac{s_nS}{WT}}-1\big) x^{\alpha}$ into \eqref{eqn:laplace_transform_inf_m}, we have
\small{\begin{align}
p_{n,j,d_j}(s_n,x)=\exp\left(-\frac{2\pi}{\alpha}\lambda_b\big(2^{\frac{s_nS}{WT}}-1\big)^{\frac{2}{\alpha}}x^2
B'\left(\frac{2}{\alpha},1-\frac{2}{\alpha},2^{-\frac{s_nS}{WT}}\right)\right).\notag
\end{align}}\normalsize
Next, we calculate the probability of successfully decoding the signal from BS $j$ after successfully decoding and canceling the signals from the nearer BSs
in $\{1,2,\cdots j-1\}$, denoted as $p_{n,j}(s_n)$, by removing the condition  of $p_{n,j,d_j}(s_n,x)$ on $d_j=x$. Note that we have the p.d.f. of
$d_j$ as $f_{d_j}(x)=\frac{2\pi^j\lambda_b^jx^{2j-1}}{(j-1)!}\exp(-\pi\lambda_bx^2)$. Thus, we have
\small{\begin{align}
p_{n,j}(s_n)&\triangleq\int_0^{\infty}p_{n,j,d_j}(s_n,x)f_{d_j}(x){\rm d}x\notag\\
&=\int_{0}^{\infty}\exp\left(-\frac{2\pi}{\alpha}\lambda_b\big(2^{\frac{s_nS}{WT}}-1\big)^{\frac{2}{\alpha}}x^2
B'\left(\frac{2}{\alpha},1-\frac{2}{\alpha},2^{-\frac{s_nS}{WT}}\right)\right)\frac{2\pi^j\lambda_b^jx^{2j-1}}{(j-1)!}\exp(-\pi\lambda_bx^2){\rm d}x\notag\\
&=\frac{1}{\left(1+\frac{2}{\alpha}\big(2^{\frac{s_nS}{WT}}-1\big)^{\frac{2}{\alpha}}B'\left(\frac{2}{\alpha},1-\frac{2}{\alpha},2^{-\frac{s_nS}{WT}}\right)\right)^j}.\notag
\end{align}}\normalsize
Finally, by assuming the independence between the events $W{\rm log}(1+{\rm SIR}_j)>\frac{s_nS}{T},\ j\in\{1,2,\cdots,i\}$\cite{Quek14TCOM}, we have
\small{\begin{align}
h(s_n,i)=\prod_{j=1}^{i}p_{n,j}(s_n)=\frac{1}{\left(1+\frac{2}{\alpha}\big(2^{\frac{s_nS}{WT}}-1\big)^{\frac{2}{\alpha}}
B'\left(\frac{2}{\alpha},1-\frac{2}{\alpha},2^{-\frac{s_nS}{WT}}\right)\right)^{\frac{i(i+1)}{2}}},\quad s_n\in\mathcal S\setminus\{0\}\;.\notag
\end{align}}\normalsize
We complete the proof of Lemma~\ref{Lem:exp_of_h_s_i}.

\section*{Appendix B: Proof of Lemma~\ref{Lem:CP_asymp_0}}

Consider any file $n\in\mathcal N$.
When $s_n=0$, we have $q^c_{n}(s_n)=0$ as $S\to 0$.
It remains to calculate $q^c_n(s_n)$ as $S\to 0$ for $s_n\in \mathcal S\setminus\{0\}$.
We note that $B^{'}(a,b,z)=\frac{(1-z)^{b}}{b}+o\left((1-z)^{b}\right)$, as $z\to 1$. Thus, as $S\to 0$, we have\footnote{$f(x)=o(g(x))$ means $\lim_{x\to0}\frac{f(x)}{g(x)}=0$}
\small{\begin{align}
&B^{'}\left(\frac{2}{\alpha},1-\frac{2}{\alpha},2^{-\frac{s_nS}{WT}}\right)= \frac{\left(2^{\frac{s_nS}{WT}}-1\right)^{1-\frac{2}{\alpha}}}{1-\frac{2}{\alpha}}
+o\left(\left(2^{\frac{s_nS}{WT}}-1\right)^{1-\frac{2}{\alpha}}\right).\label{eqn:Binc_low_tau}
\end{align}}\normalsize
In addition, from \eqref{eqn:h_z_i} and \eqref{eqn:q_n^c}, we have
\small{\begin{align}
q^c_n(s_n)&=h\left(s_n,\frac{1}{s_n}\right)=\frac{1}{\left(1+\frac{2}{\alpha}\big(2^{\frac{s_nS}{WT}}-1\big)^{\frac{2}{\alpha}}
B'\left(\frac{2}{\alpha},1-\frac{2}{\alpha},2^{-\frac{s_nS}{WT}}\right)\right)^{\frac{s_n+1}{2s_n^2}}}\notag\\
&\eqla \frac{1}{\left(1+\frac{2}{\alpha-2}\big(2^{\frac{s_nS}{WT}}-1\big)
+o\big(2^{\frac{s_nS}{WT}}-1\big)\right)^{\frac{s_n+1}{2s_n^2}}}
\eqlb \frac{1}{\left(1+\frac{2}{\alpha-2}\frac{({\rm ln}2) s_nS}{WT}
+o\big(\frac{({\rm ln}2) s_nS}{WT}\big)\right)^{\frac{s_n+1}{2s_n^2}}}\notag\\
&\eqlc \left(1-\frac{2}{\alpha-2}\frac{ ({\rm ln}2)s_nS}{WT}
+o\left(S\right)\right)^{\frac{s_n+1}{2s_n^2}}
\eqld 1-\frac{2}{\alpha-2}\frac{({\rm ln}2) s_nS}{WT}\frac{s_n+1}{2s_n^2}
+o\big(S\big)\notag
\end{align}
\begin{align}
&= 1-\frac{{\rm ln}2}{\alpha-2}\frac{ S}{WT}\left(1+\frac{1}{s_n}\right)
+o\big(S\big),\quad {\rm as }\ S\to 0, \label{eqn:q_n_asymp_0_exact}
\end{align}}\normalsize
where $(a)$ is due to \eqref{eqn:Binc_low_tau},  $(b)$ is due to $2^{x}-1=({\rm ln}2)x+o(x)$ as $x\to 0$, $(c)$ is due to
$\frac{1}{1+x}=1-x+o(x)$ as $x\to 0$, and $(d)$ is duo to the series expension $(1-x)^b=1-bx+o(x)$ at $x=0$.
Therefore, we have
\small{\begin{align}
q^c(\mathbf s)&=\sum_{n\in\mathcal N}a_n\left(1-\frac{{\rm ln}2}{\alpha-2}\frac{S}{WT}\left(1+\frac{1}{s_n}\right)+o(S)\right)\mathbf 1\left[s_n\neq 0\right] \notag\\
&=\sum_{n\in\mathcal N}a_n\mathbf 1\left[s_n\neq 0\right]-\frac{{\rm ln}2}{(\alpha-2)WT}S \sum_{n\in\mathcal N}a_n\left(1+\frac{1}{s_n}\right)\mathbf 1\left[s_n\neq 0\right]+o(S),\quad {\rm as }\ S\to 0.\notag
\end{align}}\normalsize
Thus, we have $\lim_{S\to 0}\frac{q^c(\mathbf s)}{q^c_0(\mathbf s)}=1$ implying  $q^c(\mathbf s)\stackrel{S\to0}{\sim}q^c_{0}(\mathbf s)$. Therefore, we complete the proof of Lemma \ref{Lem:CP_asymp_0}.

\section*{Appendix C: Proof of Lemma~\ref{Lem:CP_asymp_CP}}

Consider any file $n\in\mathcal N$.
When $s_n= 0$, we have $q^c_{n}(s_n)=0$ as $S\to \infty$.
It remains to calculate $q^c_n(s_n)$ as $S\to \infty$ for $s_n\in \mathcal S\setminus\{0\}$.
We note that $B'\left(a,b,z\right)=B(a,b)-\frac{z^a}{a}+o(z^a)$, as $z\to 0$. Thus, as $S\to \infty$, we have
\small{\begin{align}
&B'\left(\frac{2}{\alpha},1-\frac{2}{\alpha},2^{-\frac{s_nS}{WT}}\right)=B\left(\frac{2}{\alpha},1-\frac{2}{\alpha}\right)-
\frac{\alpha}{2}\left(2^{\frac{s_nS}{WT}}-1\right)^{-\frac{2}{\alpha}}+o\left(\left(2^{\frac{s_nS}{WT}}-1\right)^{-\frac{2}{\alpha}}\right).\label{eqn:Binc_high_tau}
\end{align}}\normalsize
In addition, from \eqref{eqn:h_z_i} and \eqref{eqn:q_n^c}, we have
\small{\begin{align}
q^c_n(s_n)=h\left(s_n,\frac{1}{s_n}\right)=& \frac{1}{\left(1+\frac{2}{\alpha}\big(2^{\frac{s_nS}{WT}}-1\big)^{\frac{2}{\alpha}}
B'\left(\frac{2}{\alpha},1-\frac{2}{\alpha},2^{-\frac{s_nS}{WT}}\right)\right)^{\frac{s_n+1}{2s_n^2}}}\notag\\
\eqla& \frac{1}{\left(\frac{2}{\alpha}\big(2^{\frac{s_nS}{WT}}-1\big)^{\frac{2}{\alpha}}
B\left(\frac{2}{\alpha},1-\frac{2}{\alpha}\right)+o(1)\right)^{\frac{s_n+1}{2s_n^2}}}\notag\\
=& \frac{1}{\left(\frac{2}{\alpha}B\left(\frac{2}{\alpha},1-\frac{2}{\alpha}\right)\big(2^{\frac{s_nS}{WT}}-1\big)^{\frac{2}{\alpha}}\right)^{\frac{s_n+1}{2s_n^2}}}
\frac{1}{\left(1+o\left(\big(2^{\frac{s_nS}{WT}}-1\big)^{-\frac{2}{\alpha}}\right)\right)^{\frac{s_n+1}{2s_n^2}}}\notag\\
=& \frac{1}{\left(\frac{2}{\alpha}B\left(\frac{2}{\alpha},1-\frac{2}{\alpha}\right)2^{\frac{s_nS}{WT}\frac{2}{\alpha}}\right)^{\frac{s_n+1}{2s_n^2}}}
\frac{1}{\left(1-2^{-\frac{s_nS}{WT}}\right)^{\frac{s_n+1}{\alpha s_n^2}}}
\frac{1}{\left(1+o\left(2^{-\frac{s_nS}{WT}\frac{2}{\alpha}}\right)\right)^{\frac{s_n+1}{2s_n^2}}}\notag\\
\eqlb& \frac{2^{-\frac{S}{WT}\frac{s_n+1}{\alpha s_n}}}{\left(\frac{2}{\alpha}B\left(\frac{2}{\alpha},1-\frac{2}{\alpha}\right)\right)^{\frac{s_n+1}{2s_n^2}}}
\left(1+\frac{s_n+1}{\alpha s_n^2}2^{-\frac{s_nS}{WT}}+o\left(2^{-\frac{s_nS}{WT}}\right)\right)
\left(1-o\left(2^{-\frac{s_nS}{WT}\frac{2}{\alpha}}\right)\right)\notag\\
=&\frac{2^{-\frac{(s_n+1)S}{\alpha s_nWT}}}{\left(\frac{2}{\alpha}B\left(\frac{2}{\alpha},1-\frac{2}{\alpha}\right)\right)^{\frac{s_n+1}{2s_n^2}}}+o\left(2^{-\frac{(s_n+1)S}{\alpha s_nWT}}\right), \quad {\rm as }\ S\to \infty,\label{eqn:q_n_asymp_inf_exact}
\end{align}}\normalsize
where $(a)$ is due to \eqref{eqn:Binc_high_tau}, and $(b)$ is due to $\frac{1}{(1+x)^b}=1-bx+o(x)$, $\frac{1}{(1-x)^b}=1+bx+o(x)$.
Therefore,  we have
\small{\begin{align}
q^c(\mathbf s)=&\sum_{n\in\mathcal N}a_n\left(\frac{2^{-\frac{(s_n+1)S}{\alpha s_nWT}}}{\left(\frac{2}{\alpha}B\left(\frac{2}{\alpha},1-\frac{2}{\alpha}\right)\right)^{\frac{s_n+1}{2s_n^2}}}+o\left(2^{-\frac{(s_n+1)S}{\alpha s_nWT}}\right)\right)\mathbf 1[s_n\neq 0]\notag
\end{align}
\begin{align}
=&\frac{2^{-\frac{(s_{\rm max}+1)S}{\alpha s_{\rm max}WT}}}{\left(\frac{2}{\alpha}B\left(\frac{2}{\alpha},1-\frac{2}{\alpha}\right)\right)^{\frac{s_{\rm max}+1}{2s_{\rm max}^2}}}\sum_{n\in\mathcal N}a_n\mathbf 1\left[s_n=s_{\rm max}\right]+o\left(2^{-\frac{(s_{\rm max}+1)S}{\alpha s_{\rm max}WT}}\right), \quad {\rm as }\ S\to \infty.\notag
\end{align}}\normalsize
Thus, we have $\lim_{S\to \infty}\frac{q^c(\mathbf s)}{q^c_{\infty}(\mathbf s)}=1$ implying $q^c(\mathbf s)\stackrel{S\to \infty}{\sim}q^c_{\infty}(\mathbf s)$. Therefore, we complete the proof of Lemma \ref{Lem:CP_asymp_CP}.

\section*{Appendix D: Proof of Lemma~\ref{Lem:opt_asymp_0}}

Substituting feasible solution $\mathbf s^*$ given in \eqref{eqn:opt_s_asymp_0} into \eqref{eqn:CP_coding_caching_aympt_0}, we have
\small{\begin{align}
q^c_0(\mathbf s^*) = \sum_{n\in\mathcal N}a_n \mathbf 1[s_n^*\neq 0]-\frac{({\rm ln}2)(1+M)S}{(\alpha-2)WT}\sum_{n\in\mathcal N}a_n \mathbf 1[s_n^*\neq 0].\notag
\end{align}}\normalsize
On the other hand, for any feasible solution $\mathbf s$ to Problem~\ref{prob:opt_coding_asymp_0},
we have
\small{\begin{align}
q^c_0(\mathbf s)=\sum_{n\in\mathcal N}a_n \mathbf 1[s_n\neq 0]-\frac{({\rm ln}2)S}{(\alpha-2)WT}\sum_{n\in\mathcal N}a_n\left(1+\frac{1}{s_n}\right)\mathbf 1[s_n\neq 0].\notag
\end{align}}\normalsize
Thus, we have
\small{\begin{align}
q^c_0(\mathbf s^*)-q^c_0(\mathbf s)=&\bigg(\sum_{n\in\mathcal N}a_n\mathbf 1[s_n^*\neq 0]-\sum_{n\in\mathcal N}a_n\mathbf 1[s_n\neq 0]\bigg)\notag\\
&-\frac{({\rm ln}2)(1+M)S}{(\alpha-2)WT}\bigg(\sum_{n\in\mathcal N}a_n\mathbf 1[s_n^*\neq 0]-\sum_{n\in\mathcal N}a_n\frac{1+\frac{1}{s_n}}{1+M}\mathbf 1[s_n\neq 0]\bigg).\label{eqn:minus_apdix_D}
\end{align}}\normalsize
Note that under any feasible solution to Problem~\ref{prob:opt_coding_asymp_0}, $u_0$ can obtain at most $KM$ files from the $M$ nearest BSs.
Under feasible solution $\mathbf s^*$,  $u_0$ can obtain the $KM$ most popular files from the $M$ nearest BSs. Thus, for any feasible solution $\mathbf s\neq \mathbf s^*$, we have
\small{\begin{align}
\sum_{n\in\mathcal N}a_n\mathbf 1[s_n^*\neq 0]>\sum_{n\in\mathcal N}a_n\mathbf 1[s_n\neq 0].\label{eqn:ineq_1}
\end{align}}\normalsize
On the other hand, if $s_n\in\mathcal S\setminus\{0\}$, we have $\frac{1}{s_n}\leq M$, implying $\frac{\frac{1}{s_n}+1}{M+1}\leq 1$. Thus,  we have
\small{\begin{align}
\sum_{n\in\mathcal N}a_n\mathbf 1[s_n\neq 0]\geq \sum_{n\in\mathcal N}a_n\frac{1+\frac{1}{s_n}}{1+M}\mathbf 1[s_n\neq 0].\label{eqn:ineq_2}
\end{align}}\normalsize
By \eqref{eqn:ineq_1} and \eqref{eqn:ineq_2},  we have
\small{\begin{align}
\sum_{n\in\mathcal N}a_n\mathbf 1[s_n^*\neq 0]> \sum_{n\in\mathcal N}a_n\frac{1+\frac{1}{s_n}}{1+M}\mathbf 1[s_n\neq 0].\label{eqn:ineq_3}
\end{align}}\normalsize
Based on \eqref{eqn:minus_apdix_D}, we have
\small{\begin{align}
&q^c_0(\mathbf s^*)-q^c_0(\mathbf s)>0\notag\\
\Longrightarrow& \ \frac{({\rm ln}2)(1+M)S}{(\alpha-2)WT}\bigg(\sum_{n\in\mathcal N}a_n\mathbf 1[s_n^*\neq 0]-\sum_{n\in\mathcal N}a_n\frac{1+\frac{1}{s_n}}{1+M}\mathbf 1[s_n\neq 0]\bigg) < \sum_{n\in\mathcal N}a_n\mathbf 1[s_n^*\neq 0]-\sum_{n\in\mathcal N}a_n\mathbf 1[s_n\neq 0]\notag\\
\stackrel{(a)}{\Longrightarrow}&\ S<\frac{\sum_{n\in\mathcal N}a_n\mathbf 1[s_n^*\neq 0]-\sum_{n\in\mathcal N}a_n\mathbf 1[s_n\neq 0]}
{\frac{({\rm ln}2)(1+M)}{(\alpha-2)WT}\left(\sum_{n\in\mathcal N}a_n\mathbf 1[s_n^*\neq 0]-\sum_{n\in\mathcal N}a_n\frac{1+\frac{1}{s_n}}{1+M}\mathbf 1[s_n\neq 0]\right)}
\triangleq g_0(\mathbf s),\notag
\end{align}}\normalsize
where $(a)$ is due to  \eqref{eqn:ineq_3}.
By \eqref{eqn:ineq_1} and \eqref{eqn:ineq_3}, we know $g_0(\mathbf s)>0$ for any feasible solution $\mathbf s \neq \mathbf s^*$.
Thus, we have $S_0\triangleq\min\left\{g_0(\mathbf s) \big{|}\; \mathbf s\neq\mathbf s^*,
\eqref{eqn:cache_size_constr_coding},\eqref{eqn:division_choise_req}\right\}>0$.
%
Therefore, when $S<S_0$, for any feasible solution $\mathbf s\neq\mathbf s^*$, we have $q^c_0(\mathbf s^*)-q^c_0(\mathbf s)>0$. We complete the proof of Lemma~\ref{Lem:opt_asymp_0}.

\section*{Appendix E: Proof of Lemma~\ref{Lem:opt_asymp_infty}}

Substituting feasible solution $\mathbf s^*$ given in \eqref{eqn:opt_s_asymp_inf} into \eqref{eqn:CP_coding_caching_aympt}, we have
\small{\begin{align}
q^c_{\infty}(\mathbf s^*) = \frac{\sum_{n\in\mathcal N}a_n\mathbf 1[s_n^*=1]}{\frac{2}{\alpha}B\left(\frac{2}{\alpha},1-\frac{2}{\alpha}\right)}2^{-\frac{2S}{\alpha WT}}>0.\notag
\end{align}}\normalsize
On the other hand, for any feasible solution $\mathbf s$ to Problem~\ref{prob:opt_coding_asymp_infty},
we have
\small{\begin{align}
q^c_{\infty}(\mathbf s)=\frac{\sum_{n\in\mathcal N}a_n\mathbf 1[s_n=s_{\rm max}]}{\left(\frac{2}{\alpha}B\left(\frac{2}{\alpha},1-\frac{2}{\alpha}\right)\right)^{\frac{s_{\rm max}+1}{2s_{\rm max}^2}}}2^{-\frac{S(s_{\rm max}+1)}{\alpha WTs_{\rm max}}}.\notag
\end{align}}\normalsize
Thus, we have
\small{\begin{align}
\frac{q^c_{\infty}(\mathbf s)}{q^c_{\infty}(\mathbf s^*)}=\frac{\sum_{n\in\mathcal N}a_n\mathbf 1[s_n=s_{\rm max}]}{\sum_{n\in\mathcal N}a_n\mathbf 1[s_n^*=1]}\left(\frac{2}{\alpha}B\left(\frac{2}{\alpha},1-\frac{2}{\alpha}\right)\right)^{1-\frac{s_{\rm max}+1}{2s_{\rm max}^2}}2^{\frac{S}{\alpha WT}(1-\frac{1}{s_{\rm max}})}.\label{eqn:ratio_apdix_E}
\end{align}}\normalsize
For any feasible solution $\mathbf s\neq \mathbf s^*$ to Problem~\ref{prob:opt_coding_asymp_infty}, we show $\frac{q^c_{\infty}(\mathbf s)}{q^c_{\infty}(\mathbf s^*)}<1$ for large $S$ by considering the following  three cases.
(i) If $s_{\rm max}=1$, based on \eqref{eqn:ratio_apdix_E}, we have
\small{\begin{align}
\frac{q^c_{\infty}(\mathbf s)}{q^c_{\infty}(\mathbf s^*)}=\frac{\sum_{n\in\mathcal N}a_n\mathbf 1[s_n=1]}{\sum_{n\in\mathcal N}a_n\mathbf 1[s_n^*=1]}.\label{eqn:ratio_short_apdix_E}
\end{align}}\normalsize
Note that for any feasible solution to Problem~\ref{prob:opt_coding_asymp_0}, $u_0$ can  obtain at most $K$ files from the nearest BSs. For feasible solution $\mathbf s^*$,  $u_0$ can obtain the $K$ most popular files from the nearest BSs. Thus, for any feasible solution $\mathbf s\neq \mathbf s^*$, we have
\small{\begin{align}
\sum_{n\in\mathcal N}a_n\mathbf 1[s_n^*=1]>\sum_{n\in\mathcal N}a_n\mathbf 1[s_n=1].\label{eqn:ineq_1_apdix}
\end{align}}\normalsize
By \eqref{eqn:ratio_short_apdix_E} and \eqref{eqn:ineq_1_apdix}, we have $\frac{q^c_{\infty}(\mathbf s)}{q^c_{\infty}(\mathbf s^*)}<1$.
(ii) If $s_{\rm max}\in\left\{\frac{1}{2},\frac{1}{3},\cdots,\frac{1}{M}\right\}$, we have
\small{\begin{align}
&\frac{q^c_{\infty}(\mathbf s)}{q^c_{\infty}(\mathbf s^*)}<1
\Longrightarrow  S>\frac{\alpha WT}{\frac{1}{s_{\rm max}}-1}{\rm log}_2\left(\frac{\sum_{n\in\mathcal N}a_n\mathbf 1[s_n=s_{\rm max}]}{\sum_{n\in\mathcal N}a_n\mathbf 1[s_n^*=1]}\left(\frac{2}{\alpha}
B\left(\frac{2}{\alpha},1-\frac{2}{\alpha}\right)\right)^{1-\frac{s_{\rm max}+1}{2s_{\rm max}^2}}\right)\triangleq g_{\infty}(\mathbf s).\notag
\end{align}}\normalsize
Define $S_{\infty}\triangleq \max\left\{\max\bigg\{g_{\infty}(\mathbf s)\Big{|} s_{\rm max}\in\left\{\frac{1}{2},\frac{1}{3},\cdots,\frac{1}{M}\right\},
\eqref{eqn:cache_size_constr_coding},\eqref{eqn:division_choise_req}\bigg\},0\right\}$. Therefore, when $S>S_{\infty}$, for any feasible solution $\mathbf s\neq\mathbf s^*$,
we have $\frac{q^c_{\infty}(\mathbf s)}{q^c_{\infty}(\mathbf s^*)}<1$.
We complete the proof of Lemma~\ref{Lem:opt_asymp_infty}.



\end{document}